\documentclass[a4paper,fleqn]{cas-sc}

\usepackage[authoryear,longnamesfirst]{natbib}
\usepackage[most]{tcolorbox} 
\usepackage{url}
\newcommand{\manuscriptstatus}{Manuscript accepted to \textit{Computers in Human Behavior}}
\ExplSyntaxOn
\cs_set:Npn \__first_footerline:
  {\group_begin:\small\sffamily\manuscriptstatus\group_end:}
\ExplSyntaxOff

\lstdefinestyle{promptstyle}{
    basicstyle=\ttfamily\small,
    breaklines=true,
    breakatwhitespace=true,
    columns=fullflexible,
    keepspaces=true
}
\newtcblisting{promptbox}[1][]{
    enhanced,
    breakable,
    colback=gray!5,
    colframe=gray!60,
    fonttitle=\bfseries\small,
    listing only,
    listing options={style=promptstyle},
    title={#1},
    boxrule=0.5pt,
    arc=2pt,
    left=6pt,
    right=6pt,
    top=4pt,
    bottom=4pt,
    toptitle=3pt,
    bottomtitle=3pt,
}

\def\tsc#1{\csdef{#1}{\textsc{\lowercase{#1}}\xspace}}
\tsc{WGM}
\tsc{QE}

\begin{document}
\let\WriteBookmarks\relax
\def\floatpagepagefraction{1}
\def\textpagefraction{.001}

\shorttitle{AI-Mediated Communication Can Reduce Opinion-Based Network Segregation}

\shortauthors{Huq et al.}

\title [mode = title]{AI-Mediated Communication Can Reduce Opinion-Based Network Segregation}  

%

\author[1]{Faria Huq}
\author[1]{Elijah L. Claggett}
\author[1]{Hirokazu Shirado\corref{cor1}}
\affiliation[1]{organization={Kenneth C. Griffin School of Computer Science, Carnegie Mellon University},
                addressline={5000 Forbes Avenue},
                city={Pittsburgh},
                state={PA},
                postcode={15213},
                country={USA}}
\cortext[cor1]{Corresponding author}
\ead{shirado@cmu.edu}



\begin{abstract}
Everyday communication is increasingly AI-mediated, with large language models (LLMs) suggesting wording, tone, and framing while humans decide whether to adopt them. 
Because communication can shape who people choose to interact with, the effects of AI mediation may extend beyond individual attitudes and perceptions to broader patterns of network segregation across opinion differences. 
Here, we examine this possibility through a preregistered online experiment (\textit{N}=711 across 80 sessions) in which participants discussed controversial political topics over multiple rounds and could freely reorganize into groups.
During the discussions, participants received LLM-generated message suggestions under either \textit{individual AI assistance} centered on participants’ own stances or \textit{relational AI assistance} incorporating group members’ perspectives.
We find that individual and relational AI assistance produced distinct trajectories of opinion-based network segregation.
Segregation remained largely unchanged under individual AI assistance but decreased significantly under relational AI assistance.
Even though many participants adopted few or no AI-generated suggestions, relational AI assistance significantly increased positive sentiment in group discussions.
More positive conversations then helped sustain more opinion-diverse groups.
The relational communication strategy had no detectable network-level effect when presented as direct instruction to participants, suggesting that AI scaffolding contributed to the observed reduction in segregation.
Incorporating the perspectives of social partners into AI-mediated communication can reduce opinion-based network segregation.

\end{abstract}



\begin{keywords}
AI-mediated communication \sep Large language models \sep Social networks \sep Polarization \sep Social sorting
\end{keywords}

\maketitle

\section{Introduction}

When people participate in online discussions, they often choose to interact with others who share their views \citep{waller2021quantifying, cinelli2021echo, Barbera_Jost_Nagler_Tucker_Bonneau_2015, gonzalez2023asymmetric}.
Over time, these repeated choices can produce broader patterns of opinion-based network segregation, a network structure in which interactions occur disproportionately among people with similar opinions \citep{mcpherson2001birds, centola2007homophily, newman2003mixing}. 
Such segregation can limit the exchange of information across diverse groups \citep{granovetter1973strength} and create environments in which misinformation, stereotypes, and prejudice are more likely to spread \citep{starbird2019disinformation, Stein_2023, perdue1990us, cinelli2021echo}.
Importantly, the way people communicate can influence whom they choose to interact with in the future \citep{yeomans2020conversational, falkenberg2024patterns}.
Communication can therefore affect whether social networks become more or less segregated over time.

One important recent change in communication is the growing role of artificial intelligence (AI) mediation \citep{hancock2020ai}.
AI tools powered by large language models (LLMs) can suggest what to say and how to say it, influencing everyday conversations.
Research shows that these tools can affect people's attitudes toward conversational partners \citep{hohenstein2020moral, williams2026biased} and help people communicate in ways that make others more willing to listen and engage \citep{argyle2023leveraging, claggett2025relational, tessler2024ai}.
Recent work further shows that AI exposure can influence subsequent human--human communication beyond direct human--AI interactions \citep{yakura2024empirical, riedl2026cognitive}. 
We know much less, however, about whether and how AI-mediated communication can change who interacts with whom over time, ultimately reinforcing or mitigating opinion-based network segregation.

\begin{figure}
    \centering
    \includegraphics[width=0.8\linewidth]{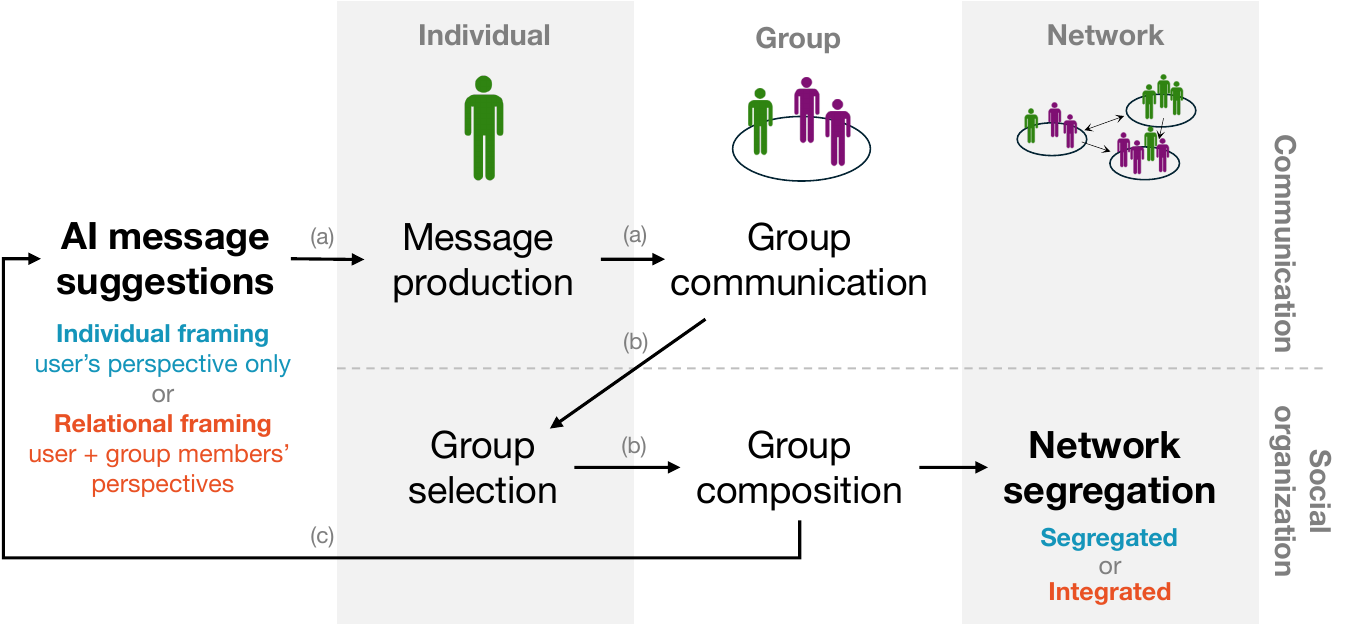}
    \caption{\textbf{Proposed multilevel framework for how AI-mediated communication can shape opinion-based network segregation.} The framework traces how AI-mediated changes in individual communication can accumulate into changes in group composition and, ultimately, network segregation. (a) AI-generated message suggestions can influence individual message production and group communication. We examine two complementary forms of AI assistance: individual framing, which incorporates only the user's perspective, and relational framing, which incorporates both the user's and the group members' perspectives. (b) Group communication can subsequently influence each individual's group selection, altering group composition. Across the entire network, these changes determine whether interactions become more segregated or integrated with respect to opinion differences over time. (c) Moreover, AI assistance may incorporate the evolving group composition into subsequent message suggestions, creating a recursive link between communication and social organization. Colored labels indicate the contrasting outcomes hypothesized under individual and relational AI framing.}
    \label{fig:mechanism}
\end{figure}

We hypothesize that AI-mediated communication can shape opinion-based network segregation through a series of connected social processes (Fig. \ref{fig:mechanism}).
AI message suggestions influence how people communicate during group discussions.
These conversations, in turn, can influence whether people stay in their current groups, leave them, or join new ones.
As people select among groups over time, group composition changes, gradually altering the broader network.
Thus, the direction of these network effects may depend on how AI message suggestions are framed at the outset.

Many existing AI communication tools are designed around individual users, helping them express their own intentions more effectively \citep{buschek2021impact, gero2023social, jakesch2023co}. 
We refer to this individual-focused approach as \textit{individual AI assistance}.
When people with different opinions engage in discussion, such assistance may emphasize users’ own perspectives and make their communication less receptive to different views \citep{minson2020won}. 
If this makes constructive interaction across disagreement harder to sustain, people may become more likely to sort into groups whose members share similar opinions, reinforcing segregation.

An alternative is for AI to help people communicate by considering not only what they want to say but also the perspectives of the people they are talking to \citep{claggett2025relational}.
We call this relationship-focused approach \textit{relational AI assistance}.
Relational AI assistance need not change people’s underlying beliefs; instead, it may help them communicate more constructively across disagreement \citep{giles1975speech, yeomans2020conversational}.
This may make continued interaction across differing viewpoints easier to sustain, encouraging more diverse groups and reducing segregation.
Moreover, because relational AI assistance adapts as group composition changes, it can continue supporting constructive communication as groups become more diverse.

People can adopt these communication strategies on their own, but AI may offer two practical advantages.
First, LLMs can consistently generate receptive messages across different group and conversational contexts, whereas people often struggle to do so during disagreement \citep{minson2020won, yeomans2020conversational}. 
Second, AI-powered message suggestions can increase users’ writing productivity \citep{noy2023experimental}.
Thus, participants who use AI-generated suggestions may exert greater influence on the group discussion by producing more messages than non-adopters.
As a result, AI-assisted communication strategies may become more prominent in the conversation even with only partial adoption.
These properties may make AI particularly well suited to sustaining communication changes across repeated interactions.

\begin{figure}
    \centering
    \includegraphics[width=1.0\linewidth]{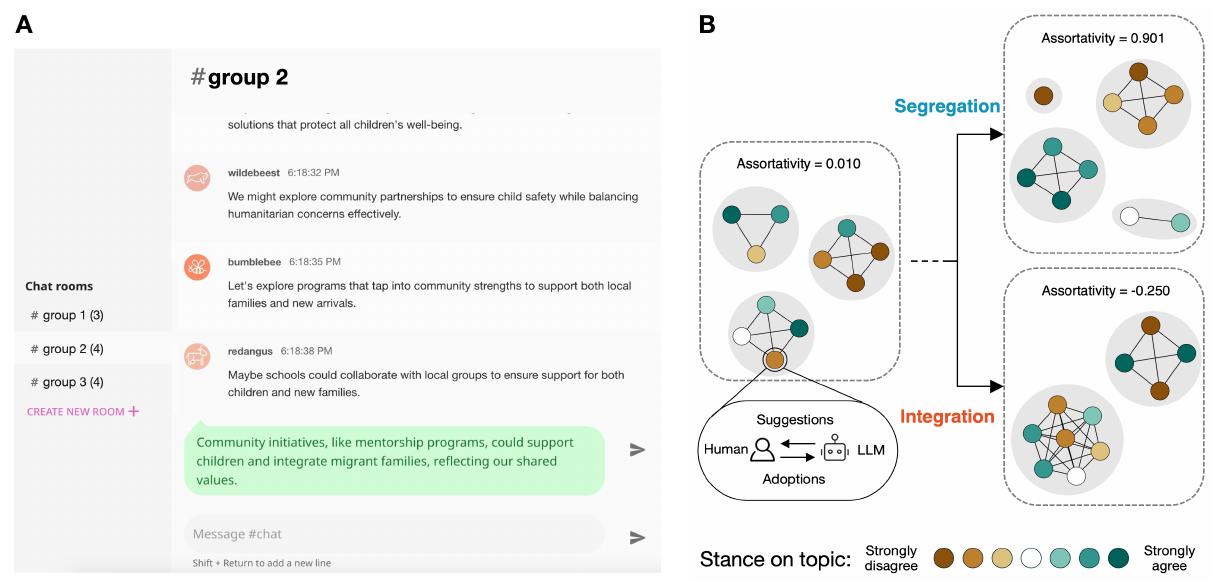}
    \caption{\textbf{Experimental interface and illustration of opinion-based network segregation.}
    (A) Participants discuss a topic in one of several chat rooms and may switch rooms over time. Numbers in parentheses indicate the number of participants currently in each room. During AI-assisted discussions, participants receive message suggestions generated by an LLM, shown as green-shaded text.
    (B) We construct the social network from discussion-group membership: two participants are connected when they belong to the same group. Nodes represent individuals, colored by opinion stance, and gray shaded regions denote discussion groups corresponding to the chat rooms in (A). As individuals move among groups over time, these connections change, and the network can become more segregated or integrated with respect to opinion differences. We quantify this pattern using assortativity by opinion stance, with higher values indicating greater segregation and lower values indicating greater integration.}
    \label{fig:networks}
\end{figure}

To examine these hypotheses, we developed an experimental system that enables participants to engage in text-based discussions across multiple groups with LLM-generated message suggestions while freely choosing which groups to join (Fig. \ref{fig:networks}).
This design captures the fluid group formation characteristic of online public discussion platforms \citep{waller2021quantifying}.
Using this system, we conducted preregistered online experiments involving 711 participants across 80 sessions \citep{HuqUnknown-wm, HuqUnknown-wm2}. 
We operationalized individual and relational AI assistance through different LLM prompting strategies and compared their effects on conversational dynamics, group formation, and ultimately network segregation.

We find that individual and relational AI assistance produce divergent conversational dynamics and distinct trajectories of  network segregation.
Without relational assistance, greater group diversity was associated with less positive communication, whereas relational assistance attenuated this tendency and helped sustain positive communication in more diverse groups over time.
Finally, we show that directly instructing participants to adopt the same relational communication strategies did not reproduce these structural effects, suggesting that AI scaffolding contributed to the observed system-level change.

In this study, we conceptualize AI-mediated communication as part of a social interaction system in which changes in communication can accumulate into changes in social organization.
Our findings show that, when AI assistance incorporates the relational context of communication, it can alter a self-reinforcing cycle toward network segregation by sustaining positive communication even as groups become more diverse.
By integrating real-time AI-mediated communication, dynamic group reorganization, and social network analysis within a controlled behavioral experiment, this work provides a multilevel experimental framework for studying AI-mediated communication in relation to broader social network dynamics.
\section{Related literature}

\subsection{AI-mediated communication}

AI-mediated communication refers to interactions in which AI systems participate in the production or transformation of human messages \citep{hancock2020ai}. 
Early examples include ``smart reply'' systems that suggest short responses to incoming text messages and emails \citep{kannan2016smart, hohenstein2020moral}. 
LLMs extend this paradigm by generating longer, context-sensitive messages and adapting communication to an ongoing interaction \citep{argyle2023leveraging, claggett2025relational}.
These systems create hybrid human--AI expression, in which communicative acts are produced jointly by users and AI systems \citep{rahwan2019machine}.

Prior work shows that AI mediation can alter both communicative behavior and interpersonal outcomes. 
In dyadic conversations, AI-generated smart replies altered participants’ linguistic patterns and increased perceived trust in their communication partners, potentially because the suggestions contained more positive emotional language \citep{hohenstein2020moral}.
In political discussions, LLM-generated rephrasings emphasizing restatement, validation, and politeness improved partners' ratings of conversation quality and democratic reciprocity without systematically shifting participants' policy positions \citep{argyle2023leveraging}. 
Moving beyond dyadic communication, \citet{tessler2024ai} showed that an AI mediator that synthesized the perspectives of small groups into shared statements increased agreement and reduced division among discussants, whereas unmediated exposure to one another’s opinions did not produce the same convergence.
Recent work has begun to show that AI influence can also spill over into subsequent human--human interaction, affecting human language production as well as shared language, collective attention, and shared mental models \citep{yakura2024empirical, riedl2026cognitive}.

However, improving communication within a conversation does not necessarily improve the broader social environment in which that conversation occurs. 
Communication may appear more positive or consensual because people have already sorted into like-minded groups, while those who feel uncomfortable, unheard, or constrained in expressing their views may leave. 
AI mediation may therefore improve the quality of ongoing interactions without necessarily increasing interaction across differences.
Prior work primarily examines how AI changes communication among preformed interaction partners; it leaves open what happens when those communication changes subsequently shape who interacts with whom.

\subsection{Communication and social structure}

Homophilous interaction can organize populations into increasingly separated social networks and communities \citep{mcpherson2001birds, waller2021quantifying}.
\citet{schelling1971dynamic} illustrates how population-level segregation can emerge from many local decisions about whom to associate with, even when individual preferences for similarity are relatively modest.
Communication can shape these local interaction dynamics because how disagreement is expressed can affect whether people remain willing to engage with one another \citep{goffman1955face, deutsch1973resolution}.
\citet{yeomans2020conversational} found that receptive language reduced subsequent conflict and made speakers more desirable partners for future collaboration.

At the same time, social structure can shape the context of subsequent communication by determining who interacts with whom and whether those interactions occur within or across social boundaries \citep{coleman1990foundations}. 
In online political communication, for example, out-group interactions are more toxic and receive lower engagement than in-group interactions, patterns that may reinforce preferences for interacting within one’s own group \citep{falkenberg2024patterns}. 
Communication and social organization may therefore evolve recursively: conversational dynamics influence subsequent interaction choices, while the resulting social structure shapes the context in which later communication occurs.

This recursive relationship is especially important for understanding political polarization and fragmentation. 
Polarization is a multilevel process that involves individual behavior and larger-scale patterns of social organization \citep{levin2021dynamics}. 
Yet these levels are often approached separately, reflecting a broader distinction between interventions that target individual behavior and those that target social systems \citep{chater2022frame, kish2024unraveling, voelkel2024megastudy}.
Recent evidence further suggests that individual-level depolarizing interventions often produce short-lived effects, underscoring the difficulty of translating local psychological change into durable societal outcomes \citep{holliday2025depolarization}. 
Understanding the effects of AI-mediated communication therefore requires considering not only its immediate effects on conversations, but also how those effects accumulate through repeated interaction choices into changes in social organization.

\subsection{From relational principles to interactional practice}

Relational approaches to communication emphasize acknowledging different perspectives and maintaining constructive relationships across differences \citep{goffman1959, emirbayer1997manifesto}.
Whereas some communication strategies focus primarily on expressing one’s own intentions, relational strategies also take the perspectives of interaction partners into account.
Yet translating these principles into effective interaction is difficult. 
Simply exposing people to opposing views does not necessarily reduce polarization and can sometimes intensify it \citep{bail2018exposure}, while people may struggle to communicate receptively despite intending to do so \citep{yeomans2020conversational}. 
These findings suggest a gap between relational principles and their practical implementation, especially during disagreement.

AI-mediated communication may help bridge this implementation gap by scaffolding relational principles at the moment of interaction \citep{claggett2025relational}.
LLMs can operationalize these principles into concrete, context-sensitive messages without requiring users to understand or formulate the relational framing themselves.
At the same time, users may choose not to adopt AI-generated suggestions, in which case the relational framing does not enter the conversation.
Moreover, AI mediation may introduce its own social costs: people may evaluate communication less positively when they believe it was generated or assisted by AI \citep{hohenstein2023artificial}, potentially because AI involvement can reduce perceived authenticity \citep{o2025unethical}.
It therefore remains unclear whether such scaffolding can sustain constructive interaction across differences and, through repeated interaction choices, alter broader patterns of social organization while preserving users’ discretion over whether to adopt AI-generated suggestions.
\section{Research questions}

Building on the gaps identified in the preceding literature, this study investigates the collective consequences of AI-mediated communication in group discussions. 
Participants discussed controversial topics in small groups while receiving no AI assistance or different forms of AI-powered message support, and could reorganize across groups over time.
Our primary research question examines how different forms of AI-mediated communication affect social network segregation (RQ1). 
We then investigate the multilevel interaction processes underlying these effects (RQ2) and the role of AI-mediated operationalization in enabling communication strategies to produce system-level structural change (RQ3).

\vspace{1.0em}

\noindent\textbf{RQ1. How do different forms of AI-mediated communication affect network segregation in opinion-diverse groups?} 
Although prior work has demonstrated individual-level effects of AI-mediated communication, its system-level consequences remain unknown—specifically, whether AI-mediated communication reinforces or mitigates network segregation.
We hypothesize that the direction of this structural effect depends on the framing of AI assistance.
AI assistance that primarily reinforces users’ own perspectives may reinforce separation between ideologically distinct groups, whereas AI assistance that incorporates and engages with others’ perspectives may reduce network segregation by encouraging more receptive communication.

\vspace{1.0em}

\noindent\textbf{RQ2. Through what multilevel interaction processes does AI-mediated communication shape network segregation?}
To explain the effects identified in RQ1, we investigate how different framings of AI assistance shape interpersonal communication dynamics, such as message sentiment and interaction patterns, which in turn influence group diversity and ultimately network segregation (Fig.~\ref{fig:mechanism}). 
These processes are expected to form recursive feedback loops, whereby emerging group composition influences subsequent communication dynamics under AI assistance.

\vspace{1.0em}

\noindent\textbf{RQ3. What role does AI-mediated operationalization play in enabling communication strategies to produce network-level change?}
To understand its role in the effects identified in RQ1, we examine whether communication strategies can produce structural change through human execution alone or whether AI-mediated operationalization contributes to sustaining their execution during repeated interaction.
To address this question, we compare AI-mediated operationalization of relational framing with equivalent direct instruction using the same relational communication strategy.

\section{Methods}

\subsection{Participants and experimental procedure}

To examine these research questions, we conducted online experiments with 711 unique participants recruited through Prolific across 80 sessions. 
The study was preregistered \citep{HuqUnknown-wm, HuqUnknown-wm2} and approved by the authors’ institutional review board. 
The main experiment was conducted in July and August 2025, followed by a supplementary experiment in May 2026. 
Participants were restricted to U.S. residents to increase the likelihood of shared familiarity with the political discussion topics. 
Table \ref{tab:demography} summarizes participant demographics.
All participants provided informed consent and received monetary compensation for completing an approximately 30-minute study.

Participants interacted anonymously through a custom online chat platform. 
Each session consisted of 6--15 participants ($M=8.89$, $SD=2.44$), who were divided into discussion groups of 3--4 people. 
Before the discussion, participants completed a short survey indicating their stances on seven controversial topics using a 7-point Likert scale (Table \ref{tab:initial_survey}). 
For each session, we calculated the Shannon entropy of participants' stance distribution for each topic and selected the topic with the highest entropy, corresponding to the issue on which participants' opinions were most evenly distributed.
In the event of a tie, one of the tied topics was selected at random.

The experiment consisted of four rounds of group discussion each lasting three minutes. 
Participants engaged in open-ended conversation with their group members (Fig. \ref{fig:networks}A) . 
Each discussion round began with a prompt related to the selected topic, such as ``\textit{Should teachers and books in public schools be banned from discussing sexuality and gender?}'' for the LGBT topic.
Depending on the experimental condition, participants either received no assistance or real-time AI-powered message suggestions while composing messages.
A separate supplementary experiment using the same procedure evaluated the instruction condition (see Sec. \ref{sec:AI}).
All messages, including AI-powered message suggestions, were recorded with timestamps by a single server.

After each discussion round, participants could remain in their current group, switch to another group, or create a new group before the next round. 
Occasionally, participants could become the sole member of a discussion group after creating a new group or after all other members switched elsewhere. 
These singleton groups arose naturally from the group-switching process; group-level analyses were restricted to groups containing at least two participants.
Participants could also update their stance during the session, although such changes occurred infrequently (6.40\% of participants changed their stance at least once).
At the end of the session, participants completed a post-task questionnaire about their experience and demographics.

Sessions in the main experiment were randomly assigned to the no-assistance, individual-assistance, or relational-assistance condition (see below). 
The supplementary experiment used the instruction condition.
Following preregistered criteria \citep{HuqUnknown-wm, HuqUnknown-wm2}, sessions with fewer than six participants or a dropout rate of 25\% or greater were excluded, resulting in a final sample of 711 participants across 80 sessions (20 sessions per condition; $N = 187$ in the no-assistance condition, $N = 182$ in the individual-assistance condition, $N = 188$ in the relational-assistance condition, and $N = 154$ in the supplementary instruction condition). 
Participant counts, dropout rates, and discussion-topic assignments did not differ significantly across conditions (participant counts: $\chi^2(3)=4.35$, $p=.226$; dropout rates: $\chi^2(3)=2.07$, $p=.559$; discussion-topic assignments: Fisher's exact test, $p=.541$).

\subsection{AI-powered message suggestions} \label{sec:AI}

In the AI assistance conditions, participants received real-time message suggestions generated by OpenAI’s \texttt{GPT-4o} model during group discussion (Fig. \ref{fig:networks}A).
A new suggestion was generated each time a new message was posted in the participant's group.
Participants could send suggestions directly, edit them before sending, or ignore them and write their own messages.

We examined two forms of AI-powered expressive support that differed in how the message suggestions were personalized.
\textit{Individual AI assistance} personalized suggestions solely according to the participant’s own viewpoint, whereas \textit{relational AI assistance} additionally incorporated the viewpoints of other group members to encourage constructive engagement. 

In the individual AI assistance condition, suggestions were generated using the participant’s own pre-task survey stance on the selected topic.
The model was instructed to ``reinforce the participant’s stance (even if it is divisive)'' and ``emphasize their perspective firmly rather than seeking compromise''.
For example, when disagreeing with stricter immigration requirements, the system could suggest: ``\textit{We need to prioritize effective resource management; fear-driven narratives only distract from implementing real, practical solutions.}'' (Table \ref{tab:sample_messages}).
This condition resembles common forms of AI personalization that aim to align with users’ preferences and intentions without incorporating broader social context \citep{cheng2026sycophantic, ibrahim2026training}.

In the relational AI assistance condition, suggestions were generated using the participant’s own stance together with the stances of the other group members \citep{claggett2025relational}. 
The model was instructed to ``acknowledge others' points, find common ground, and use a cooperative tone'' while remaining consistent with the participant’s own viewpoint.
For example, the system could suggest: ``\textit{I think prioritizing children's safety is key, but supporting migrants can enrich our community in meaningful ways too.}'' (Table \ref{tab:sample_messages}).
This condition was informed by theories of communication accommodation and relational framing \citep{giles1975speech, gumperz1982discourse, maass1989language}. 
The two conditions otherwise shared the same prompt structure and generation constraints. 
Complete prompts are provided in Appendix B.

To examine whether the observed effects were specific to AI-powered message suggestions or could be reproduced through direct guidance alone, we also conducted a supplementary instruction condition. 
In this condition, participants did not receive AI-generated suggestions. 
Instead, they were continuously shown the guidance, ``\textit{Express your opinion while acknowledging others' points, finding common ground, and using a cooperative tone.}'' in the same interface location and presentation format as the message suggestions.
The guidance was based on the same relational communication strategy used in the relational AI assistance condition (Fig. \ref{fig:interface}B).


\subsection{Measures and analysis}

Our primary measure of opinion-based network segregation was \textit{stance assortativity}, a network analogue of Pearson’s correlation coefficient that computes the correlation between the stances of participants connected through discussion-group co-membership \citep{newman2003mixing, asratian1998bipartite}.
It measures the extent to which participants are connected to others with similar stances on the discussion topic relative to what would be expected under random mixing.
Higher values indicate a greater tendency for participants with similar opinions to form groups together and thus greater segregation, whereas lower values indicate greater mixing across opinion differences (Fig. \ref{fig:networks}B). 

Preregistered session-level analyses controlled for the first-round value of each outcome to account for between-session variation. 
Consistent with successful randomization, first-round stance assortativity did not differ significantly across experimental conditions ($F(3,76)=1.33$, $p=.271$, $\eta^2=.05$). 
We then fit session-level linear regression models with stance assortativity as the outcome, controlling for its first-round value and including experimental condition, round, and their interaction as predictors, using the no-assistance condition as the reference.
The analysis included the final network structure after group switching at the end of Round 4, treating this post-switch network as Round 5.
The condition × round interaction terms estimate how the individual and the relational AI framing  altered the per-round trajectory of segregation relative to the no-assistance condition.
We applied the same analytical approach to session-level outcomes, including within-group and between-group stance distance, the number of connections, and the number of groups, as complementary structural measures underlying changes in stance assortativity.

We next explored the conversational and group-composition dynamics through which AI assistance may influence opinion-based network segregation (Fig. \ref{fig:mechanism}).
To characterize conversational dynamics, we analyzed complementary measures of communication quantity and quality.
\textit{Message volume} measures the number of messages sent by participants, whereas \textit{message sentiment} measures the overall positive or negative tone of messages.
We quantified message sentiment using the VADER sentiment analysis toolkit \citep{hutto2014vader}, where positive (negative) values indicate more positive (negative) message sentiment.
We also applied additional toolkits \citep{danescu2013computational, lees2022new} to assess two subcomponents of message positivity, namely politeness and toxicity.
These established toolkits have been validated against human annotations in prior work and provide scalable and standardized measures of message characteristics.
Taken together, these measures characterize the overall positivity of group communication, which may influence the reorganization of discussion groups and, ultimately, network segregation.
In the AI assistance conditions, we further distinguished between AI-assisted and self-written messages to examine how exposure to AI-generated language influenced participants’ own communication behavior.

We characterized group composition dynamics by examining how discussion groups reorganized over time as participants switched groups after each discussion round. 
Discussion groups were identified by their predefined names, allowing group composition to be tracked consistently across rounds.
We quantified these dynamics using within-group stance distance, defined as the average ideological distance among participants within each discussion group.
Unlike stance assortativity, which characterizes the network as a whole, within-group stance distance is a group-level measure of opinion diversity.
Because this measure is undefined for singleton groups, it was computed only for groups with at least two members.
Decreases (increases) in group-level within-group stance distance indicate increasing ideological homogeneity (heterogeneity) within discussion groups over time.

We used mixed-effects models to estimate the temporal relationships among these components while accounting for repeated observations within discussion sessions. Specifically, the models estimated (a) the effects of AI assistance conditions on individual- and group-level message volume and sentiment, (b) the effects of group-level message volume, message sentiment, and their interaction on subsequent changes in group-level within-group stance distance, and (c) whether participants' message sentiment changed as they interacted with group members holding increasingly different opinions, and whether this relationship differed across AI assistance conditions (see Table \ref{tab:model_specification} for full model specifications).

\section{Results}

\begin{figure}
    \centering
    \includegraphics[width=0.6\linewidth]{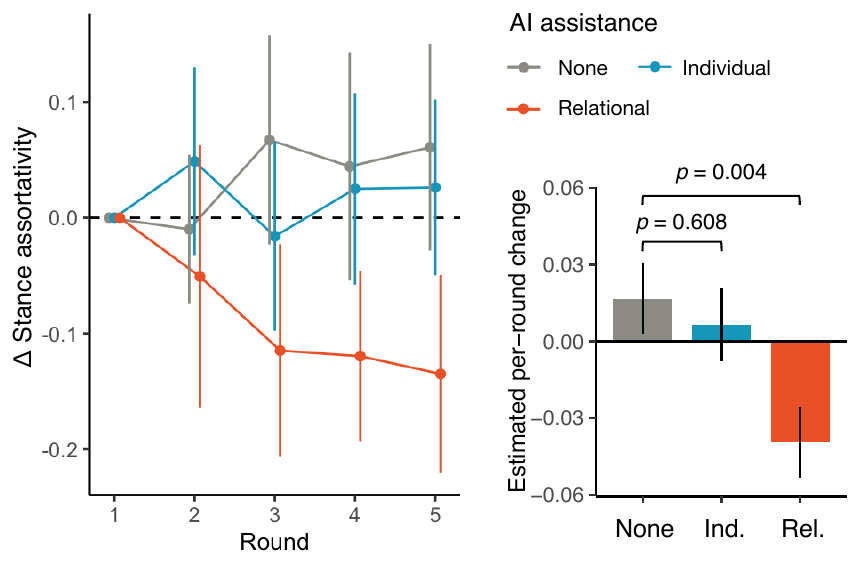}
    \caption{\textbf{Relational AI assistance reduces opinion-based network segregation.} Line plots show session-level mean changes in stance assortativity relative to the initial round, and error bars indicate ± s.e.m. across sessions ($N=20$ per condition).
    Relational AI assistance significantly reduced stance assortativity relative to the no-assistance condition ($p=0.004$), whereas individual AI assistance did not ($p=0.608$).
    }
    \label{fig:assortativity}
\end{figure}

\subsection{Effects of AI-powered message suggestions on opinion-based network segregation (RQ1)}

We found that stance assortativity remained stable or increased slightly across rounds in the no-assistance condition, but decreased substantially under relational assistance (Fig. \ref{fig:assortativity} and Table \ref{tab:regression_structure}). 
After four rounds, assortativity decreased by an average of 0.135 points (on the –1 to 1 scale) in the relational assistance condition, representing a significant reduction relative to the no-assistance control ($b=-0.056$, 95\% CI $[-0.095,\,-0.017]$, $p=.004$). 
In contrast, individual assistance did not significantly change assortativity relative to the control ($b=-0.010$, 95\% CI $[-0.049,\,0.029]$, $p=.608$). 
These results identify relational assistance as the only intervention that significantly reduced network segregation.

\begin{figure}
    \centering
    \includegraphics[width=0.54\linewidth]{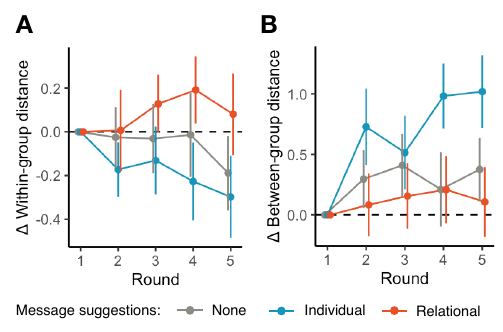}
    \caption{\textbf{Changes in within- and between-group opinion differences across AI-assistance conditions.} (A) Change from the initial round in within-group stance distance, where higher values indicate greater opinion diversity within groups. (B) Change from the initial round in between-group stance distance, where higher values indicate greater separation between groups. Error bars indicate ± s.e.m. across sessions ($N$ = 20 per condition).}
    \label{fig:structure_quality}
\end{figure}

To better understand how relational assistance reduced social network segregation, we examined within-group and between-group stance distance.
Within-group stance distance remained stable or slightly decreased in the no-assistance and individual-assistance conditions, but showed a trend toward increasing under relational assistance relative to the no-assistance condition (Fig.~\ref{fig:structure_quality}A). 
Linear regression estimated a more positive temporal slope under relational assistance ($b=0.068$, 95\% CI $[-0.006,\,0.142]$, $p=.070$).
Between-group stance distance increased significantly over discussion rounds in the no-assistance condition ($b=0.108$, 95\% CI $[0.019,\,0.197]$, $p=.017$; Fig.~\ref{fig:structure_quality}B). 
The increase was significantly greater under individual assistance ($b=0.184$, 95\% CI $[0.058,\,0.310]$, $p=.004$). 
In contrast, relational assistance did not significantly alter the temporal slope relative to the no-assistance condition ($b=-0.060$, 95\% CI $[-0.186,\,0.066]$, $p=.350$).
These patterns suggest that relational assistance maintained greater diversity within groups, whereas individual assistance was associated with increasing separation between groups.

We also examined changes in the number of conversation partners and conversation groups to further characterize these structural dynamics.
Participants increased their number of conversation partners over discussion rounds in the no-assistance condition ($b=0.358$, 95\% CI $[0.242,\,0.475]$, $p<.001$; Fig.~\ref{fig:structure_quantity}A). 
Linear regression estimated a significantly steeper increase under individual assistance than under the no-assistance condition ($b=0.185$, 95\% CI $[0.020,\,0.350]$, $p=.028$). 
In contrast, the temporal slope under relational assistance did not differ significantly from the no-assistance condition ($b=-0.031$, 95\% CI $[-0.196,\,0.134]$, $p=.712$). The total number of conversation groups remained relatively stable across conditions (Fig. \ref{fig:structure_quantity}B). 

Taken together, individual assistance increased separation between groups, whereas relational assistance reduced overall opinion-based network segregation while maintaining greater within-group diversity.

\subsection{Mechanism (RQ2)}
To understand how AI-powered message suggestions shaped social network segregation, we next examined the proposed multilevel mechanism underlying this process (Fig. \ref{fig:mechanism}).
Based on our conceptual framework, we examined three interrelated processes: (a) how AI-powered message suggestions influenced conversational dynamics, (b) how conversational dynamics influenced group composition, and (c) how group composition subsequently influenced message suggestions and conversational dynamics.

The following subsections examine each process in turn.

\subsubsection{From message suggestions to conversational dynamics}

We first examined how AI-powered message suggestions shaped communication during group discussions (process (a) in Fig.~\ref{fig:mechanism}).
Participants actively adopted AI-generated message suggestions during group discussions.
Overall, 45.1\% of messages in the individual assistance condition and 37.0\% in the relational assistance condition were AI-generated, whereas the remaining messages were written directly by participants.
Among adopted AI suggestions, 96.3\% were sent without modification. 

\begin{figure}
    \centering
    \includegraphics[width=0.54\linewidth]{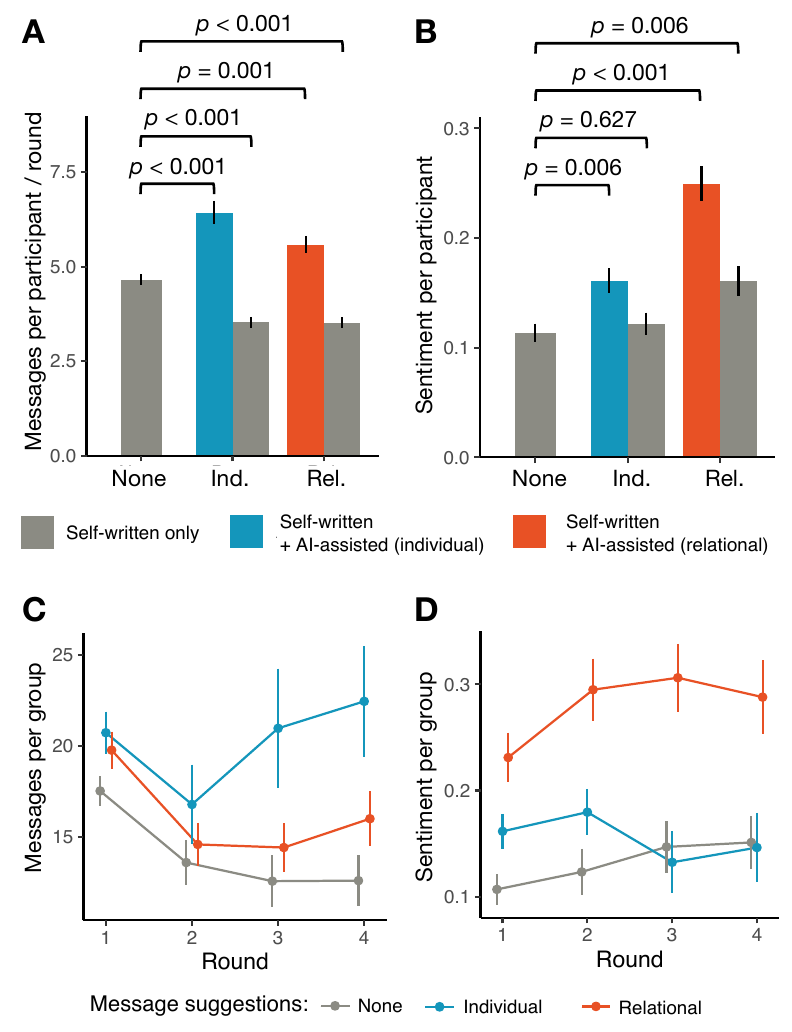}
\caption{\textbf{AI assistance differentially shapes conversational dynamics.}
(A) Individual-level message volume and (B) message sentiment. Gray bars show self-written messages only, whereas colored bars show all messages sent under the individual- and relational-assistance conditions (self-written and AI-generated messages combined). $p$-values are from regression analyses.
(C) Group-level trajectories of message volume and (D) average message sentiment over discussion rounds. Points show means and error bars indicate $\pm$ s.e.m.
}
\label{fig:message}
\end{figure}

The two forms of AI assistance substantially influenced participants’ messaging behavior (Fig. \ref{fig:message}A). 
Individual assistance significantly increased total message volume by up to 37.9\% relative to the no-assistance condition ($b = 7.11$, 95\% CI $[4.83,\, 9.38]$, $p < .001$), whereas relational assistance produced a more modest increase in message volume (up to 20.0\%; $b = 3.72$, 95\% CI $[1.47,\, 5.97]$, $p = .001$). 
At the same time, participants wrote significantly fewer self-written messages under both individual assistance ($b = -4.50$, 95\% CI $[-6.78,\, -2.23]$, $p < .001$) and relational assistance ($b = -4.54$, 95\% CI $[-6.79,\, -2.29]$, $p < .001$) than the no-assistance condition.

The AI-generated contributions also influenced the emotional tone of participants’ messages (Fig. \ref{fig:message}B). 
As a consequence of the prompt design, relational assistance generated more emotionally positive suggestions than individual assistance (Fig. \ref{fig:vader_sentiment}). 
Participants exposed to relational assistance subsequently produced significantly more positive messages than those in the no-assistance condition ($b = 0.136$, 95\% CI $[0.102,\, 0.170]$, $p < .001$), whereas individual assistance also increased message positivity, although the effect was more modest ($b = 0.048$, 95\% CI $[0.014,\, 0.082]$, $p = .006$). 
Under relational assistance, this effect extended beyond AI-generated suggestions to participants’ own writing. 
Participants also produced more positive self-written messages than those in the no-assistance condition ($b = 0.047$, 95\% CI $[0.013,\, 0.081]$, $p = .006$). 
This increase in emotional positivity was also accompanied by greater politeness and lower toxicity (Fig. \ref{fig:convokit_perx}).

These individual-level changes scaled into distinct patterns of group-level conversational dynamics (Tables~\ref{tab:sample_discussion} and \ref{tab:regression_conversation}). 
At the group level, individual assistance produced a significantly steeper increase in message volume over discussion rounds than the no-assistance condition (mixed-effects regression, Round × Individual Assistance: $b = 2.57$, 95\% CI [0.69,\,4.45], $p = .008$; Fig. \ref{fig:message}C). 
In contrast, relational assistance produced significantly higher overall conversational sentiment ($b = 0.139$, 95\% CI [0.046,\,0.233], $p = .004$; Fig. \ref{fig:message}D). 
Individual assistance primarily amplified participation, whereas relational assistance shifted both AI-assisted and self-written communication toward more positive interactions.

\begin{figure}
  \centering
  \includegraphics[width=0.54\linewidth]{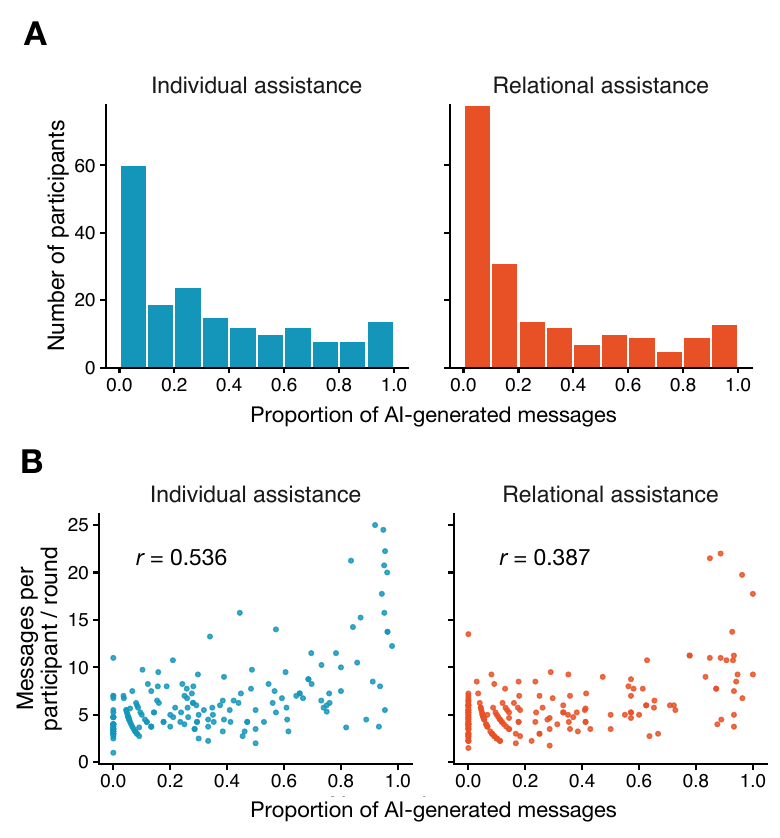}
    \caption{\textbf{AI use varied substantially across participants and was associated with message production.}
    (A) Histograms show the proportion of each participant's messages that were AI-generated in the individual-assistance ($N=182$) and relational-assistance ($N=188$) conditions. Zero indicates that all messages were self-written, whereas one indicates that all messages were AI-generated. AI use varied substantially across participants in both conditions.
    (B) Participants with higher proportions of AI-generated messages sent more messages under both assistance conditions. $r$ indicates Spearman's rank correlation; both correlations are significant at $p<0.001$.}
  \label{fig:AI_use_histogram}
\end{figure}

Notably, AI usage was unevenly distributed across participants (Fig. \ref{fig:AI_use_histogram}A).
More than 70\% of participants used AI-generated suggestions for fewer than half of their messages (71.4\% under individual assistance and 75.5\% under relational assistance), while 18.1\% and 25.5\%, respectively, did not use any AI-generated suggestions.
At the same time, individuals who relied more heavily on AI suggestions contributed more messages to group discussions under both individual and relational assistance (Spearman’s $r = .536$ and $r = .387$, respectively; both $p < .001$; Fig. \ref{fig:AI_use_histogram}B).
Thus, participants who relied more heavily on AI accounted for a disproportionate share of AI-mediated communication, making it possible for individual or relational framing to become prominent in group conversations despite limited adoption by many participants.

\subsubsection{From conversational dynamics to group diversity}
\begin{figure}
    \centering
    \includegraphics[width=0.9\linewidth]{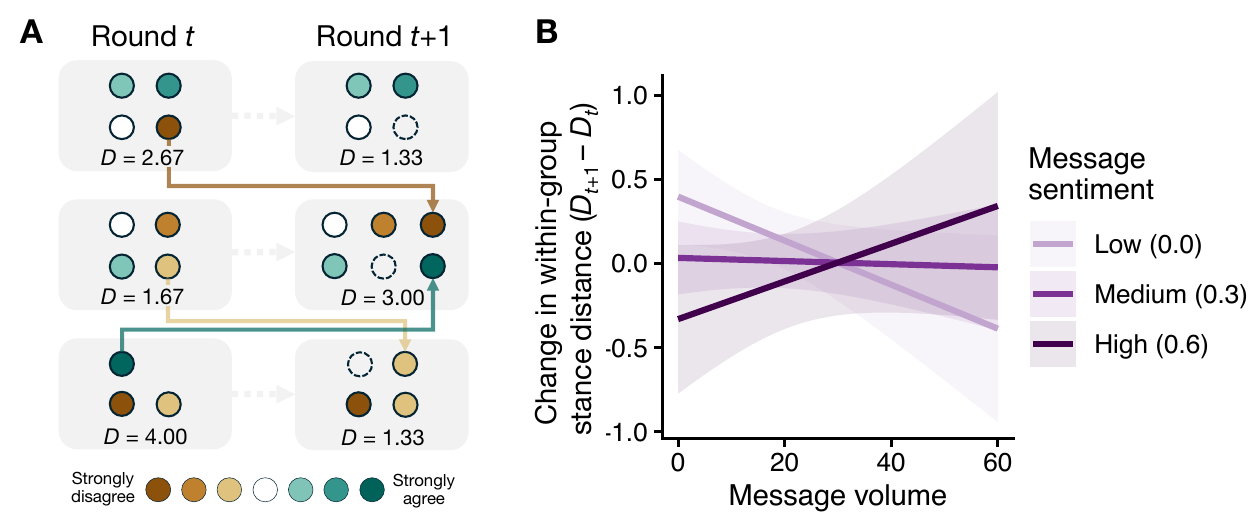}
    \caption{\textbf{Effects of conversational dynamics on group diversity.} 
(A) Illustration of how group switching changes within-group stance distance across consecutive discussion rounds. From Round~$t$ to Round~$t{+}1$, three participants switch groups, resulting in changes in within-group stance distance $D$. 
(B) Predicted change in within-group stance distance from Round~$t$ to Round~$t{+}1$ as a function of group-level message volume and message sentiment, estimated using a linear mixed-effects model (Table \ref{tab:regression_change}). Lines represent model predictions evaluated at message sentiment scores of 0.0, 0.3, and 0.6 for illustration. Prior within-group stance distance, group size, and round are held constant at their median values. Shaded regions indicate 95\% confidence intervals.}
    \label{fig:diversity_change}
\end{figure}

We next examined how conversational dynamics shaped subsequent changes in group composition (process (b) in Fig.~\ref{fig:mechanism}).
After each discussion round, an average of 32.6\% of participants left their groups to join another group or create a new one. 
Using the no-assistance condition as the reference category, a mixed-effects logistic regression showed no significant difference in the likelihood of switching in either the individual-assistance condition (OR = 1.13, 95\% CI [0.66, 1.95], p = .658) or the relational-assistance condition (OR = 1.14, 95\% CI [0.66, 1.97], p = .626).
While participants became progressively less likely to switch groups over discussion rounds (OR = 0.465, 95\% CI $[0.380,\, 0.565]$, $p<.001$; Fig. \ref{fig:group_switching}), this temporal trend did not differ in the AI-assisted conditions relative to the no-assistance condition (both interactions $ps>.90$).
Thus, the structural differences observed in Fig.~\ref{fig:assortativity} did not arise from differences in how often participants switched groups, but rather from how they reorganized across groups (Fig. \ref{fig:diversity_change}A).

We then tested whether group-level conversational dynamics predicted subsequent changes in group diversity. 
Specifically, we examined whether group-level message volume, message sentiment, and their interaction in a given round predicted changes in within-group stance distance in the following round, controlling for the group’s current stance distance. 
Figure \ref{fig:diversity_change}B shows a significant interaction between message volume and message sentiment ($b = 0.042$, 95\% CI $[0.001,\,0.082] $, $p = .043$; Table~\ref{tab:regression_change}). 
This interaction indicates that exchanging a larger number of positively framed messages was associated with greater subsequent increases in within-group stance diversity.
Because participants rarely changed their own stances (6.40\% of cases), changes in within-group stance diversity primarily reflect shifts in group composition rather than ideological movement by individuals.

These changes in group diversity were not confined to participants who heavily relied on AI assistance. 
We found no significant relationship between participants’ AI usage rates and changes in their stance distance over the course of the session (Spearman’s $r = -0.050$ and $r = 0.098$ for the individual and relational assistance conditions, respectively; both $p > .100$; Fig. \ref{fig:AI_use_diversity}). 
This finding suggests that the observed structural changes emerged through group-level interaction dynamics rather than simply reflecting the behavior of participants who used AI most frequently.

Because the proposed mechanism spans multiple levels—from individual communication to group interactions and ultimately emergent patterns of network segregation (Fig. \ref{fig:mechanism})—we complement the preceding group-level analyses with a session-level path analysis examining AI assistance condition, conversational dynamics, and subsequent stance assortativity (Fig. \ref{fig:path}; Table \ref{tab:sem_estimates}).
Consistent with the proposed mechanism, the effects of AI-powered message suggestions on subsequent stance assortativity were statistically accounted for by conversational dynamics.
In particular, relational assistance substantially increased message sentiment, whereas neither AI assistance condition showed a significant direct effect on subsequent stance assortativity after accounting for conversational dynamics.

\subsubsection{From group diversity to subsequent communication}
Finally, we examined whether increasing group diversity subsequently shaped participants' communication through AI-generated message suggestions (process (c) in Fig.~\ref{fig:mechanism}).
Unlike individual assistance, relational assistance explicitly incorporated the perspectives of current group members when generating message suggestions, maintaining positive sentiment in its suggestions even as disagreement within groups increased. 
Accordingly, relational assistance might help maintain positive communication even as participants became increasingly distant in opinion stance from their group members.

To examine this recursive effect, we fit a linear mixed-effects regression predicting participants' message sentiment from their mean stance distance from other group members, AI assistance condition, and their interaction (Table \ref{tab:regression_feedback_sentiment}). 
In the no-assistance condition, participants’ message sentiment became less positive as their mean stance distance from other group members increased ($b = -0.028$, 95\% CI $[-0.046,\,-0.010]$, $p = .002$). 
This negative relationship was significantly attenuated under relational assistance (interaction: $b=0.026$, 95\% CI $[0.001,0.050]$, $p=.045$), resulting in little decline in message sentiment as stance distance increased (Fig.~\ref{fig:feedback}). 
The interaction under individual assistance was similar in direction but did not reach statistical significance ($b=0.021$, 95\% CI $[-0.005,0.047]$, $p=.114$).
A corresponding group-level analysis yielded the same pattern (Table \ref{tab:regression_feedback_sentiment_group}).

These results indicate that relational AI assistance altered how participants communicated as they interacted with increasingly opinion-diverse groups.
Rather than exhibiting the decline in message positivity observed in the no-assistance condition, participants with relational AI assistance maintained similarly positive communication even when interacting with others holding more dissimilar stances. 
These findings are consistent with the proposed feedback mechanism, whereby relational assistance sustains positive communication as groups become more opinion-diverse, contributing to the sustained group diversity and lower network segregation observed over time.

\subsection{AI-powered message suggestions vs. direct instruction (RQ3)}

To determine whether the observed reduction in opinion-based network segregation could be reproduced by relational framing alone, we tested whether direct instruction could reproduce the effects of relational AI-powered message suggestions.
In a supplementary condition, participants received the same relational framing strategy through persistent on-screen instructions during group discussions, but without AI-generated message suggestions (Fig. \ref{fig:interface}B).

\begin{figure}
    \centering
    \includegraphics[width=0.65\linewidth]{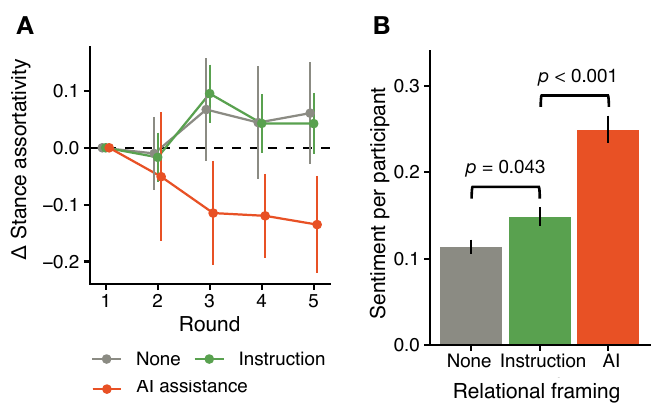}
    \caption{\textbf{Direct instruction did not reproduce the network-level effect of relational AI assistance.} 
    (A) Changes in stance assortativity across discussion rounds under no assistance, relational instruction, and relational AI-powered message suggestions. Values are shown relative to the first round. Error bars indicate $\pm$ s.e.m. 
    (B) Average message sentiment per participant across conditions. Error bars indicate $\pm$ s.e.m.; $p$-values are from regression analyses.
    Results for the no assistance and relational AI assistance conditions are identical to Figures \ref{fig:assortativity} and \ref{fig:message}B and are shown for comparison.}
    \label{fig:supp}
\end{figure}

We found that direct instruction did not reproduce the effects of AI-powered message suggestions on network segregation (Fig. \ref{fig:supp}A). 
Under direct instruction, stance assortativity remained relatively stable across rounds, closely resembling the dynamics observed in the no-assistance condition ($b = 0.0005$, 95\% CI $[-0.025,\,0.026]$, $p = .970$). 
In contrast, relational AI-powered message suggestions produced a significantly greater reduction in stance assortativity over time than direct instruction ($b = -0.041$, 95\% CI $[-0.067,\,-0.015]$, $p = .002$), despite both conditions using the same relational framing strategy. 
These results indicate that explicit relational instructions alone were insufficient to produce the system-level structural changes observed in Fig. \ref{fig:assortativity}.

Analysis of message sentiment further clarifies the distinction between individual-level behavioral change and system-level structural change (Fig. \ref{fig:supp}B). 
While direct instruction significantly increased message sentiment relative to the no-assistance condition ($b = 0.036$, 95\% CI $[0.001,\, 0.071]$, $p = .043$), the resulting sentiment was significantly lower than that under AI-powered message suggestions ($b = 0.100$, 95\% CI $[0.065,\, 0.135]$, $p < .001$). 
This smaller behavioral effect was not accompanied by measurable changes in network segregation (Fig. \ref{fig:supp}A). 
This pattern suggests that system-level structural changes may require sustained communication scaffolding beyond direct instruction—a capability well suited to contemporary AI-mediated communication.

\section{Discussion}

\subsection{AI-mediated communication shapes network segregation through multilevel dynamics}

This study conceptualizes AI-mediated communication not merely as a tool for individual expression, but as a component of a broader social interaction system.
Relational AI assistance reduced network segregation by sustaining more positive communication in opinion-diverse groups, whereas individual AI assistance increased message production without producing comparable structural effects.
These effects emerged through reciprocal dynamics between communication and group organization: conversational patterns shaped how participants reorganized across groups, while emerging group composition subsequently shaped communication under AI assistance.
Based on these findings, we discuss three theoretical implications for understanding AI-mediated communication at the level of social systems.

First, improvements in individual communication do not necessarily lead to less segregated social networks.
Individual AI assistance increased message production but also widened separation between groups, whereas relational assistance reduced segregation by sustaining interaction across opinion differences (Fig. \ref{fig:assortativity}).
AI may encourage people to communicate more, but that does not necessarily mean they will engage more with people who disagree with them.
These structural effects do not necessarily require fundamentally new strategies of communication. 
AI may simply make certain communication strategies easier and more consistent by reducing the effort needed to produce messages.
Nevertheless, over repeated interactions, these relatively small changes in communication can accumulate and reshape the broader social network.

Second, AI-mediated communication and social structure influence each other over time, consistent with classic theories of the reciprocal relationship between social structure and individual action \citep{giddens1984constitution, coleman1990foundations}.
As participants reorganized into groups, changes in group composition shaped subsequent communication.
In the no-assistance condition, communication became less positive as opinion differences within groups increased.
Relational AI assistance substantially attenuated this tendency by incorporating the perspectives of current group members into message suggestions.
These dynamics suggest that communication and social organization can evolve recursively over time.
The supplementary experiment further showed that direct relational instruction improved conversational sentiment but did not reproduce the structural effect of relational AI assistance. 
This suggests that providing AI-generated message suggestions during ongoing interactions may be more effective in shaping social network structure than simply instructing people to adopt the same communication strategies.

Third, the consequences of AI-mediated communication in groups depend on who contributes to the interaction, not only on how each individual's communication changes.
Group conversations often exhibit substantial inequality in participation \citep{woolley2010evidence, nonnecke2000lurker}, and participation patterns can shape group-level outcomes \citep{berger1980status, luskin2022deliberative}.
AI use was likewise highly uneven in our experiment, and participants who relied more heavily on AI suggestions sent more messages. 
The structural consequences of AI assistance may not require widespread adoption if frequent AI users contribute disproportionately to group interaction \citep{shirado2017locally}.
This also suggests that individual-level adoption rates and behavioral effects should not be assumed to scale directly to the group- and system-level impacts of AI-mediated communication.

\subsection{Design and social implications for AI-mediated communication}

Many existing AI personalization approaches optimize for individual users' preferences and intentions \citep{cheng2026sycophantic, ibrahim2026training}. 
Although such personalization can improve individual productivity \citep{noy2023experimental}, our findings suggest that optimizing communication for individual users may also shape how groups form and separate over time.
This motivates an alternative paradigm—\textit{relational personalization}—in which AI assistance also considers the social context in which communication occurs \citep{claggett2025relational}. 
Whereas conventional personalization can be framed as “\textit{Help me say what I want},” relational personalization asks “\textit{Help me say what I want to the people I am talking to}.”
The distinction is therefore not between helping users express their views and constraining those views, but between optimizing communication primarily for the individual and helping that individual communicate within a social context.
This shift in the objective of personalization may influence both individual communication and the broader social structures.

A second design implication concerns how communication strategies are operationalized.
Communication interventions can seek to increase awareness of constructive communication strategies \citep{balietti2021reducing}, whereas AI message suggestions can help users enact such strategies repeatedly during ongoing interaction \citep{mckee2023scaffolding, hohenstein2023artificial}. 
Because individual-level communication changes may weaken as they propagate toward collective outcomes, sustained system-level effects may require interventions that repeatedly scaffold communication as interactions unfold  \citep{ruggeri2024synthesis}. 

AI assistants may be well suited to this role because they can provide persistent, context-sensitive support throughout ongoing social interaction.
Their advantage over direct instruction may arise from several complementary forms of scaffolding: reducing the effort required to formulate messages, translating abstract communication principles into concrete language, adapting suggestions to the current conversational context, and providing support repeatedly at the moment of interaction.
Our experiment does not isolate the contribution of each of these features, but the comparison with direct instruction suggests that communicating the relational principle alone is insufficient to reproduce the same structural effects.

These findings suggest a pathway that connects interpersonal and system-level approaches to polarization and collective fragmentation \citep{finkel2020political, waller2021quantifying, cinelli2021echo}. 
Prior work has often treated interventions aimed at receptiveness and interventions aimed at exposure structure as separate strategies \citep{roozenbeek2023countering, hartman2022interventions}. 
Relational AI links the two levels by shaping interpersonal communication in ways that can alter subsequent social organization.
Rather than exogenously modifying who people encounter \citep{bakshy2015exposure, bail2018exposure} or seeking immediate consensus within a preformed group \citep{tessler2024ai}, relational AI may support the endogenous formation of cross-cutting social connections.
Such endogenous connections may help sustain communication across disagreement while preserving autonomy in social ties and viewpoints \citep{granovetter1973strength, baldassarri2020diversity}.

At the same time, relational personalization raises important design and governance questions. 
Incorporating other group members’ perspectives may improve audience awareness, but it requires systems to infer, represent, and act on socially sensitive information. 
Designing such systems requires attention to transparency, consent, privacy, and the possibility that automated efforts to promote receptiveness may oversimplify disagreement or privilege particular norms of civility.
Moreover, lower segregation is not inherently beneficial in every setting, particularly when continued interaction exposes participants to harassment, misinformation, or asymmetric power. 
AI-mediated communication may also alter perceptions of authenticity and trust in public discourse, where participants expect to engage with one another’s own words rather than AI-assisted expression \citep{o2025unethical}. 
Relational AI should therefore be understood not as a general prescription for increasing interaction across differences, but as a means of sustaining cross-cutting engagement in settings where continued interaction is itself desirable.

\subsection{Limitations and future work}

Although random assignment allowed us to identify the causal effect of AI-powered message suggestions on opinion-based network segregation, the analyses of the intervening mechanisms were not themselves experimentally controlled.
The observed associations among communication patterns, group-switching decisions, and evolving network structure may therefore remain subject to residual confounding, selection effects, and dynamic reciprocal relationships.
For example, group-switching decisions may have reflected interface affordances, participants' understanding of the switching process, or other practical considerations not captured by our measures.
Future work should combine targeted experimental manipulations of these proposed mechanisms with models that more directly capture the dynamic dependencies among communication, group reorganization, and network structure.

Second, our analyses relied on automated language measures to characterize sentiment and related conversational properties.
These tools may imperfectly capture context-dependent, culturally variable, or strategically expressed language, introducing measurement error into the estimated communication pathways.
Because the study relied primarily on quantitative behavioral and automated language measures, it also provides limited insight into how participants interpreted AI suggestions or why they chose to remain in or leave particular groups.
Qualitative coding, interviews, or post-task elicitation could help distinguish the psychological and practical processes underlying the observed dynamics.

Third, our experimental paradigm allowed participants to freely reorganize into new discussion groups, making changes in group composition a central mechanism through which communication influenced social structure.
This feature was designed to capture the fluid group formation observed on online public discussion platforms \citep{waller2021quantifying}.
In many real-world settings, however, social relationships are constrained by workplaces, families, schools, or other enduring institutions \citep{berger1966social}.
Under such conditions, AI-mediated communication may reshape social structure through different behavioral responses, such as changes in engagement, participation, or migration across affiliated groups and communities.
Future work should examine whether similar interaction dynamics emerge in these more constrained social environments.

Fourth, AI adoption was highly uneven across participants and differed across assistance conditions.
Participants used relational AI suggestions less frequently than individual AI suggestions, raising an additional question about what drives selective adoption.
This difference may reflect users' preferences—for example, they may be more willing to adopt suggestions that reinforce their existing views—or properties of the generated suggestions themselves, such as their perceived relevance, naturalness, or quality.
Future work could experimentally vary adoption rates and usage intensity, directly manipulate the effort required to enact different communication strategies, and examine why users selectively adopt particular forms of AI-mediated expression.

Fifth, the design space of AI-powered message suggestions remains large.
Our study examined only a limited set of expressive framing strategies centered on relational receptiveness.
Future work could investigate how suggestion quality, timing, controllability, transparency, memory, and the amount of message-production work performed by AI affect both adoption and downstream collective outcomes \citep{bai2025llm,kobis2025delegation, ibrahim2026training}.

Finally, our study examined AI-mediated communication in relatively small groups interacting synchronously over a short time horizon.
This controlled setting enabled us to isolate the causal effects of AI-powered message suggestions despite the complex feedback dynamics linking communication and social network evolution.
However, many real-world communication environments involve much larger populations, asynchronous interactions, longer-term participation, and broader institutional and algorithmic influences \citep{avalle2024persistent,ibrahim2026systematic}.
The effectiveness of relational AI may also differ across cultural contexts, communication settings, and domains involving different forms of social identity, conflict, or misinformation \citep{starbird2019disinformation,falkenberg2022growing}.
Future work should examine whether similar interaction dynamics emerge in these more complex settings and how AI-mediated communication influences social structure over extended periods of time.

\section{Conclusion}

This study demonstrates that AI-mediated communication can reduce social network segregation in opinion-diverse groups. 
AI-powered message suggestions with relationship-focused framing fostered more positive communication, reduced network segregation, and sustained greater diversity within evolving discussion groups. 
These structural changes emerged through recursive interactions linking communication patterns, group reorganization, and social network evolution over time.
Our findings further suggest that sustained AI scaffolding is important for translating individual communication strategies into collective structural change.
Direct instruction alone did not achieve the same reduction in opinion-based network segregation observed with relationship-focused AI message suggestions, highlighting the potential importance of sustained communication scaffolding in translating changes in individual communication into broader social network outcomes.

As AI-mediated communication becomes increasingly integrated into everyday social interaction, understanding how AI shapes the emergence of social structure will become increasingly important for the design of future communication platforms and collective interaction systems.
AI assistant systems could be designed not only to optimize individual users’ immediate goals, but also to support the long-term social relationships and collective structures that emerge through repeated interaction.




\printcredits

\section*{Acknowledgments}
We thank R. E. Kraut, S. Rathje, and N. A. Christakis for their valuable feedback on the manuscript.
This research was supported by the National Science Foundation (Grant No. 2237095) and the NOMIS Foundation.

\section*{Declaration of generative AI and AI-assisted technologies in the manuscript preparation process}

During the preparation of this work, the author(s) used OpenAI's ChatGPT to assist with language refinement and editing of the manuscript. All generated content was carefully reviewed and revised by the authors as needed. The authors take full responsibility for the content of the published article.

\section*{Data and code availability}

The processed data and analysis code supporting the
findings of this study are publicly available on Zenodo
at \url{https://doi.org/10.5281/zenodo.24492491}.

\bibliographystyle{cas-model2-names}
\bibliography{papers-ref}

\appendix
\clearpage
\newpage
\section{Appendix}

\renewcommand{\thefigure}{A\arabic{figure}}

\renewcommand{\thetable}{A\arabic{table}}

\setcounter{figure}{0}

\setcounter{table}{0}

\subsection{Additional Figures and Tables}\label{secA1}
\clearpage
\begin{figure}
    \centering
    \includegraphics[width=0.9\linewidth]{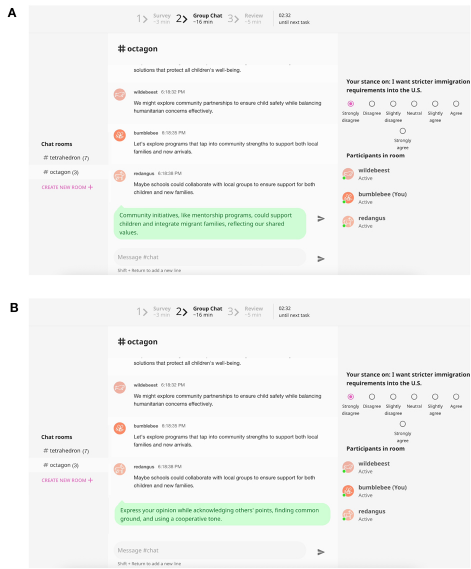}
    \caption{\textbf{Experimental discussion interface with (A) AI-powered message suggestions and (B) direct relational framing instruction.} 
    (A) Participants discussed controversial political topics in small groups while receiving AI-generated message suggestions. The interface displayed the ongoing group conversation, the participant’s stance on the topic, current group members, and a suggested message generated by the AI system incorporating relational context. Participants could freely adopt, edit, or ignore the suggested message while interacting with others and reorganizing into new discussion groups across rounds.
    (B) In the supplementary experiment, participants did not receive AI-generated suggestions. Instead, they received persistent on-screen instructions encouraging the same relational framing strategy used in the relational assistance condition and were introduced to this strategy during the tutorial session.}
    \label{fig:interface}
\end{figure}

\begin{figure}
    \centering
    \includegraphics[width=0.65\linewidth]{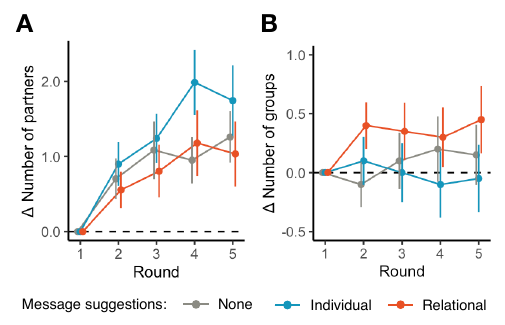}
    \caption{\textbf{Changes from the initial round in (A) number of conversation partners, and (b) number of conversation groups.} Points show means and error bars indicate ± s.e.m. 
    }
    \label{fig:structure_quantity}
\end{figure}

\begin{figure}
\centering
\includegraphics[width=0.45\linewidth]{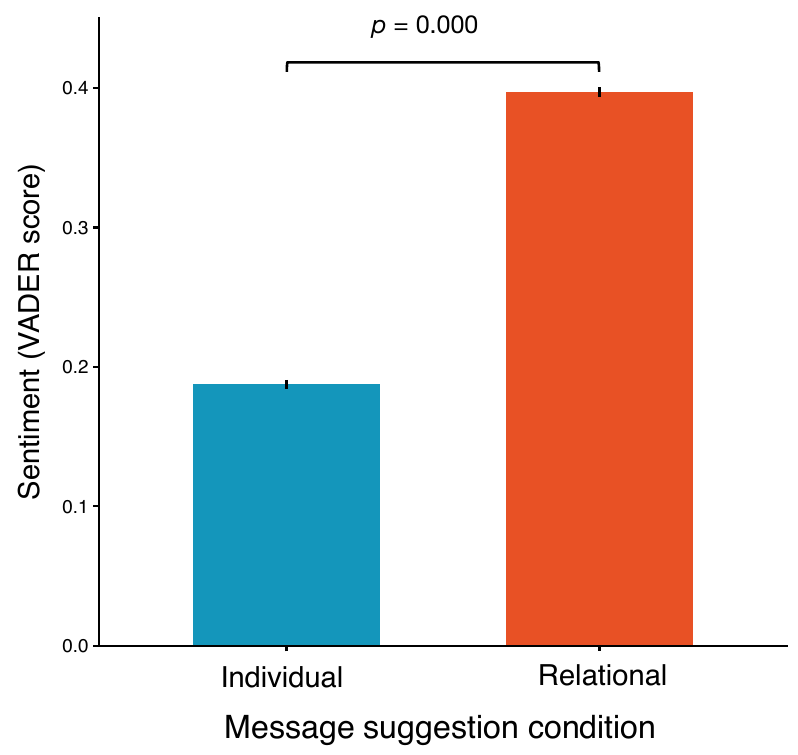}
\caption{\textbf{Sentiment of AI-provided suggestions.} Bars show the mean VADER compound sentiment score of all LLM-generated message suggestions, including those participants did not use, in the individual and relational assistance conditions. Error bars represent mean $\pm$ s.e.m.}
\label{fig:vader_sentiment}
\end{figure}

\begin{figure}
  \centering
  \includegraphics[width=0.95\linewidth]{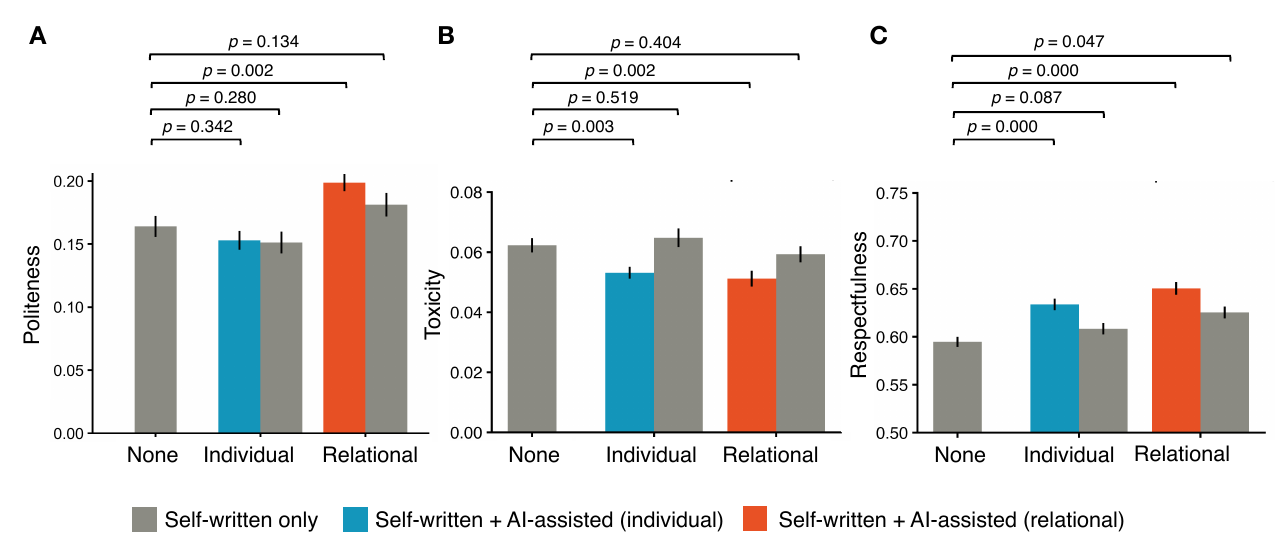}
  \caption{\textbf{Additional semantic analyses of messages across conditions.} 
  (A) Politeness, computed using the Cornell Conversation Analysis Toolkit. Context-irrelevant features (e.g., second-person usage) were removed; remaining features were min–max normalized, with negative lexicons weighted negatively. 
  (B) Toxicity (likelihood that a message could drive others away) and (C) Respectfulness (degree to which a message acknowledges others’ perspectives), both measured using the Perspective API. 
  Error bars denote mean~$\pm$~s.e.m.}
  \label{fig:convokit_perx}
\end{figure}

\begin{figure}
  \centering
  \includegraphics[width=0.6\linewidth]{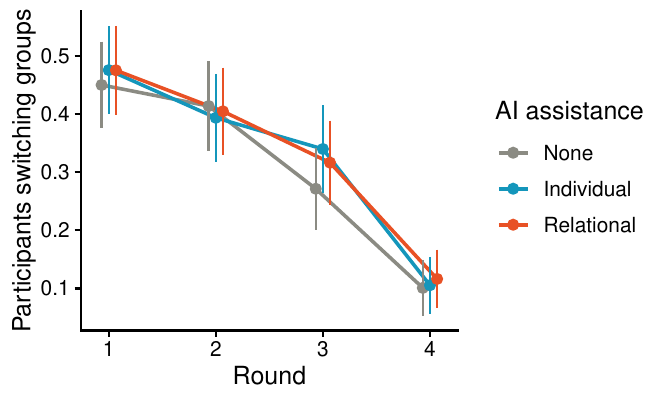}
  \caption{\textbf{Group-switching rates across rounds and AI-assistance conditions.}
  Points show the observed proportion of participants who switched groups after each round, and error bars indicate 95\% confidence intervals. A mixed-effects logistic regression showed that the likelihood of switching decreased significantly over time (OR = 0.465, 95\% CI $[0.380,\, 0.565]$, $p < 0.001$). The interactions between round and both the individual- and relational-assistance conditions were not statistically significant (both $p > 0.90$), indicating no evidence that the temporal trend differed from that in the no-assistance condition.}
  \label{fig:group_switching}
\end{figure}

\begin{figure}
  \centering
  \includegraphics[width=0.55\linewidth]{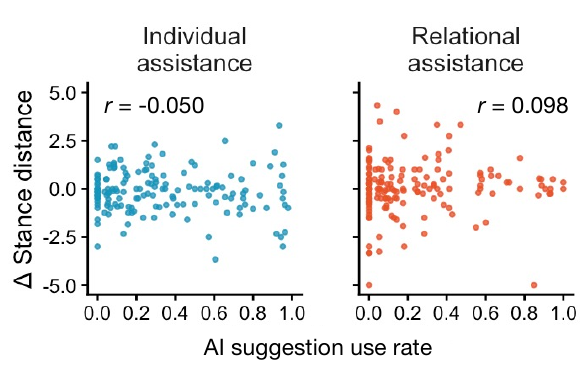}
  \caption{\textbf{Individual AI use is not associated with individual-level changes in opinion diversity.}
  Dots represent individual participants’ AI suggestion use rate and their change in average egocentric stance distance from the first to the final round under individual and relational assistance ($N=182$ in the individual-assistance condition; $N=188$ in the relational-assistance condition). Spearman’s rank correlations indicate no significant association in either condition ($r = -0.050$ for individual assistance; $r = 0.098$ for relational assistance; both $p > 0.10$), consistent with a network-level mechanism in which heterogeneous AI use does not yield corresponding individual-level diversity outcomes.}
  \label{fig:AI_use_diversity}
\end{figure}

\begin{figure}
    \centering
    \includegraphics[width=0.7\linewidth]{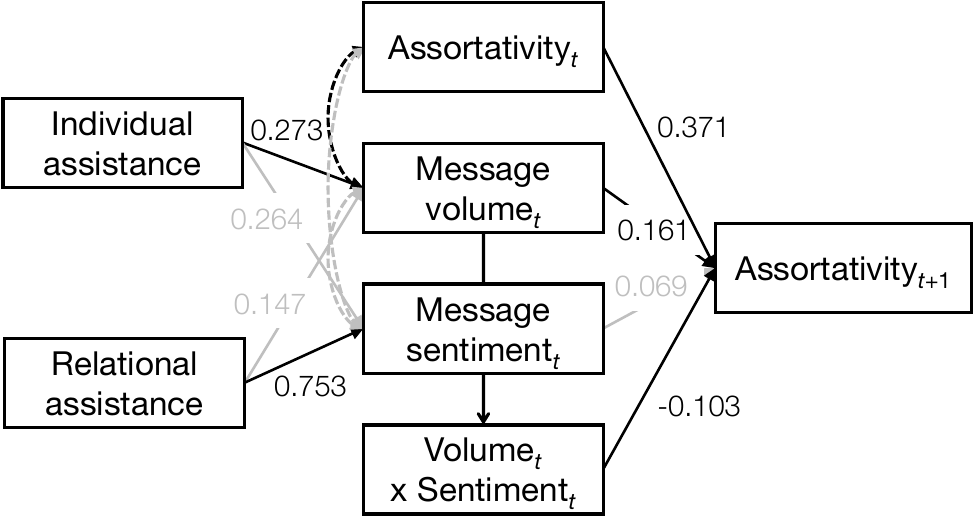}
    \caption{\textbf{Path analysis of session-level effects of AI-powered message suggestions on stance assortativity.} 
    Message volume, sentiment, and their interaction predict stance assortativity in the subsequent round, controlling for prior assortativity.
    The effects of individual and relational assistance are assessed relative to the no-assistance control condition.
    Solid arrows indicate standardized path coefficients, and dashed lines indicate covariances among mediators.
    Black lines indicate paths significant at $p < 0.001$, and gray lines indicate paths that do not meet this threshold. 
    Covariance estimates and indirect effects are provided in Table \ref{tab:sem_estimates}.}
    \label{fig:path}
\end{figure}

\begin{figure}
    \centering
    \includegraphics[width=0.5\linewidth]{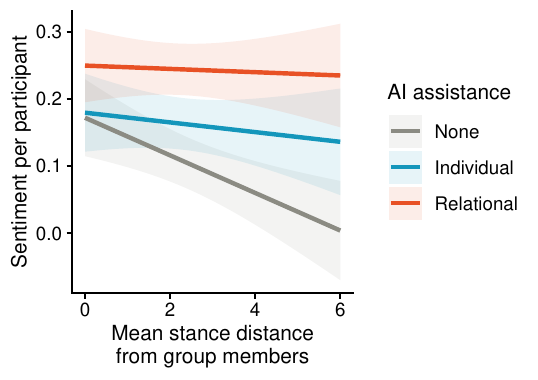}
    \caption{\textbf{Relational AI assistance maintains positive communication in opinion-diverse groups.}
    Predicted individual message sentiment as a function of participants' mean stance distance from other group members across experimental conditions, estimated using a linear mixed-effects model (Table~\ref{tab:regression_feedback_sentiment}). Lines represent model predictions and shaded regions indicate 95\% confidence intervals. Group size and round are held constant at their median values.}
    \label{fig:feedback}
\end{figure}


\clearpage

\begin{table}
    \centering
    \caption{\textbf{Demographics of experiment participants.} The data were self-reported by the participants in a post-study questionnaire ($N = 711$). }
  \label{tab:demography}
  \resizebox{0.6\textwidth}{!}{%
  \begin{tabular}{llll}
\toprule
 & Characteristics & Count & Percentage \\ \midrule
Gender & Female & 358 & 50.4\% \\
 & Male & 309 & 43.5\% \\
 & Non-binary & 15 & 2.1\% \\
 & No answer & 29 & 4.1\% \\ \midrule
Age & 18-24 & 72 & 10.1\% \\
 & 25-34 & 209 & 29.4\% \\
 & 35-44 & 196 & 27.6\% \\
 & 45-54 & 134 & 18.8\% \\
 & 55-64 & 57 & 8.0\% \\
 & 65-74 & 17 & 2.4\% \\
 & No answer & 26 & 3.7\% \\ \midrule
Ethnicity & White & 495 & 69.6\% \\
 & Black or African American & 86 & 12.1\% \\
 & Asian & 65 & 9.1\% \\
 & Hispanic or Latino & 32 & 4.5\% \\
 & Middle Eastern or North African & 5 & 0.7\% \\
 & Native American or Alaska Native & 3 & 0.4\% \\
 & No answer & 25 & 3.5\% \\ \midrule
Education & Less than high school & 4 & 0.6\% \\
 & High school diploma & 89 & 12.5\% \\
 & Some college (no degree) & 210 & 29.5\% \\ 
 & Bachelor's degree & 256 & 36.0\% \\
 & Graduate degree\textbf{} & 124 & 17.4\% \\
 & No answer & 28 & 3.9\% \\ \midrule
Annual income & \textless{}\$25,000 & 81 & 11.4\% \\
 & \$25,000-\$50,000 & 137 & 19.3\% \\
 & \$50,000-\$75,000 & 170 & 23.9\% \\
 & \$75,000-\$100,000 & 118 & 16.6\% \\
 & \textgreater{}\$100,000 & 166 & 23.3\% \\
 & No answer & 39 & 5.5\% \\ \bottomrule
\end{tabular}%
}
\end{table}

\begin{table}
\centering
\caption{\textbf{Survey items used in the initial opinion survey.} Participants rated their agreement with each statement on a seven-point Likert scale (–3 = strongly disagree, +3 = strongly agree). Discussion topics for each session were selected based on the issue showing the greatest opinion diversity among assigned participants. Numbers indicate the resulting distribution of discussion topics across experimental conditions. No significant difference was observed in the distribution ($p = 0.541$; Fisher’s exact test).}
  \label{tab:initial_survey}
  \begin{tabular}{l p{0.4\textwidth} cccc}
    \toprule
    Topic & Survey item & None & Individual & Relational & Instruction\\
    \midrule
    Evolution & I would want my kids to be taught evolution as a fact of biology. & 0 & 0 & 0 & 0\\
    Gun control & My Second Amendment right to bear arms should be protected. & 2 & 3 & 1 & 3\\
    Military spending & I support funding the military. & 4 & 2 & 3 & 5\\
    LGBT & Our children are being indoctrinated at school with LGBT messaging. & 1 & 5 & 1 & 0\\
    Climate change & I would pay higher taxes to support climate change research. & 3 & 4 & 5 & 4\\
    COVID-19 & Restrictions to stop the spread of COVID-19 went too far. & 6 & 2 & 3 & 5\\
    Immigration & I want stricter immigration requirements into the U.S. & 4 & 4 & 7 & 3\\
    \bottomrule
  \end{tabular}
\end{table}

\begin{table}
\centering
\caption{\textbf{Examples of AI message suggestions under different assistance conditions.}
The table illustrates how AI message suggestions vary across conditions and group composition.
For comparison, all examples use the same discussion topic (``I want stricter immigration requirements into the U.S.”) and the same user stance (``disagree”).
The group network shows the user`s egocentric configuration of stance differences with other group members.
Individual assistance provides suggestions that help the user express their own opinion, regardless of group composition.
In contrast, relational assistance takes the current group composition into account and adjusts the suggested messages accordingly.
When a user is surrounded by others with opposing stances, the suggestions acknowledge multiple perspectives, whereas when the user is among like-minded peers, the suggestions resemble those generated under individual assistance.}
    \includegraphics[width=0.7\linewidth]{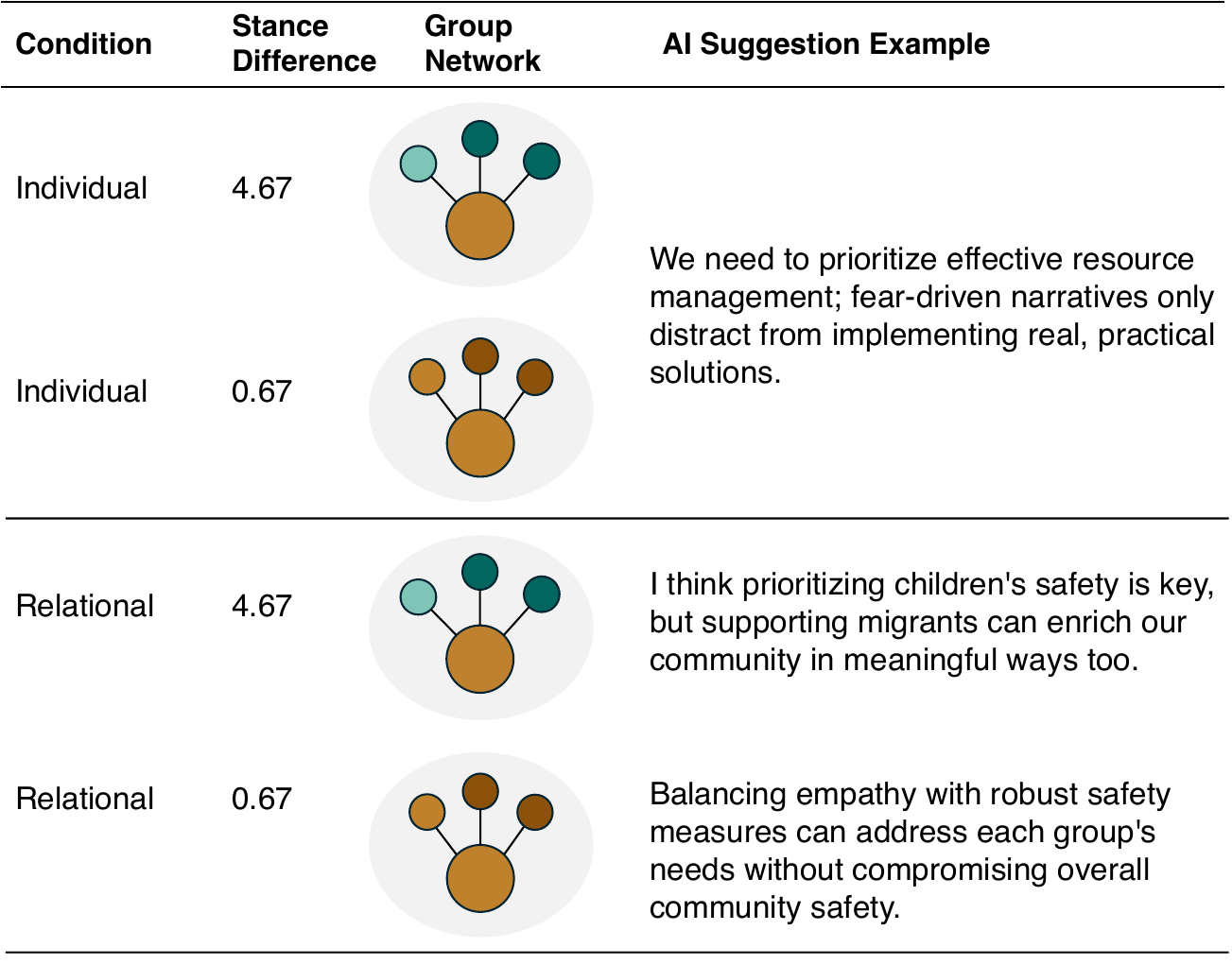}
    \label{tab:sample_messages}
\end{table}

\begin{table}[t]
\centering
\caption{\textbf{Specification of regression models used to examine the three mechanism processes.} Process labels correspond to the conceptual framework shown in Fig.~\ref{fig:mechanism}. The no-assistance condition served as the reference category in all models including AI assistance condition.}
\label{tab:model_specification}
\small
\setlength{\tabcolsep}{3pt}
\begin{tabular}{
l
>{\raggedright\arraybackslash}p{3.2cm}
>{\raggedright\arraybackslash}p{8.6cm}
>{\raggedright\arraybackslash}p{1.8cm}
}
\toprule
Process & Model & Specification & Reported in\\
\midrule

(a) &
AI $\rightarrow$ Individual messages &
Outcome: round-level average individual message volume and sentiment (VADER)& Fig. \ref{fig:message}AB \\
&& Fixed effects: AI assistance condition $\times$ message source \\
\\

&AI $\rightarrow$ Group &
Outcome: group-level message volume and sentiment (VADER) & Table \ref{tab:regression_conversation}\\
&communication& Fixed effects: AI assistance condition $\times$ round \\
&& Random intercepts: session/group\\

\midrule

(b) &
Group communication &
Outcome: next-round within-group stance distance & Table \ref{tab:regression_change}\\
&$\rightarrow$ Group diversity & Fixed effects: group-level message volume $\times$ sentiment (VADER) & \\
&& Controls: current within-group stance distance, group size, round\\
&& Random intercepts: session/group\\

\midrule

(c) &
Group diversity  &
Outcome: individual message sentiment (VADER) & Table \ref{tab:regression_feedback_sentiment} \\
&$\rightarrow$ Individual messages & Fixed effects: mean stance distance from group members $\times$ AI assistance condition & \\
&& Controls: group size, round\\
&& Random intercepts: session/group, participant\\

\bottomrule
\end{tabular}
\end{table}

\begin{table}
  \centering
  \caption{\textbf{Session-level regression results of structural measures over rounds and AI conditions ($N=300$ session-round observations).} 
  Models were estimated at the session-round level, with 60 main-experiment sessions observed over five rounds. 
  The no-assistance condition serves as the reference category. 
  Estimates represent changes per round and condition-by-round differences relative to no assistance. 
  $\dagger$ $p<0.10$, * $p<0.05$, ** $p<0.01$, *** $p<0.001$.}
  \label{tab:regression_structure}
  \begin{tabular}{llcccl}
    \toprule
    & & Estimate & 95\% CI & $P$ value &\\
    \midrule
    \textbf{Assortativity} & Round  & 0.017 & [-0.011, 0.044] & 0.230 &\\
    & Round $\times$ Individual & -0.010 & [-0.049, 0.029] & 0.608 &\\
    & Round $\times$ Relational   & -0.056 & [-0.095, -0.017] & 0.004 &** \\
    \midrule
    \textbf{Within distance} & Round & -0.029 & [-0.082, 0.023] & 0.266 &\\
    & Round $\times$ Individual & -0.047 & [-0.121, 0.026] & 0.206 &\\
    & Round $\times$ Relational   & 0.068  & [-0.006, 0.142] & 0.070 &$\dagger$ \\
    \midrule
    \textbf{Between distance} & Round & 0.108  & [0.0193, 0.197] & 0.017 &* \\
    & Round $\times$ Individual & 0.184  & [0.058, 0.310] & 0.004 &** \\
    & Round $\times$ Relational   & -0.060 & [-0.186, 0.066] & 0.350 &\\
    \midrule
    \textbf{Number of partners} &  Round  & 0.358  & [0.242, 0.475] & $<0.001$ &***\\
    & Round $\times$ Individual & 0.185  & [0.020, 0.350] & 0.028 &*\\
    & Round $\times$ Relational   & -0.031 & [-0.196, 0.134] & 0.712 &\\
    \midrule
    \textbf{Number of groups} &  Round  & 0.043  & [-0.035, 0.122] & 0.280 &\\
    & Round $\times$ Individual & -0.057  & [-0.168, 0.055] & 0.317 &\\
    & Round $\times$ Relational   & 0.083 & [-0.028, 0.195] & 0.142 &\\
    \bottomrule
  \end{tabular}
\end{table}

\begin{table}
    \centering
    \caption{\textbf{Examples of group discussions under individual and relational assistance.}
    This table presents excerpts from group discussions on U.S. immigration policy. Each message is annotated with the participant ID and their stance on the topic, based on pre-task survey responses to the item ``I want stricter immigration requirements into the U.S.” Numbers in parentheses indicate the VADER sentiment score of the message (–1 to 1; higher values indicate more positive sentiment). Messages originally generated by GPT-4o and sent by participants are labeled ``AI-assisted.”
    Under individual assistance, suggestions helped users articulate their own views but offered little support for moderating or integrating across differing positions.
    Under relational assistance, in contrast, suggestions incorporated more receptive language, helping groups maintain constructive, cross-cutting engagement despite stance differences.
    }
    \includegraphics[width=0.85\linewidth]{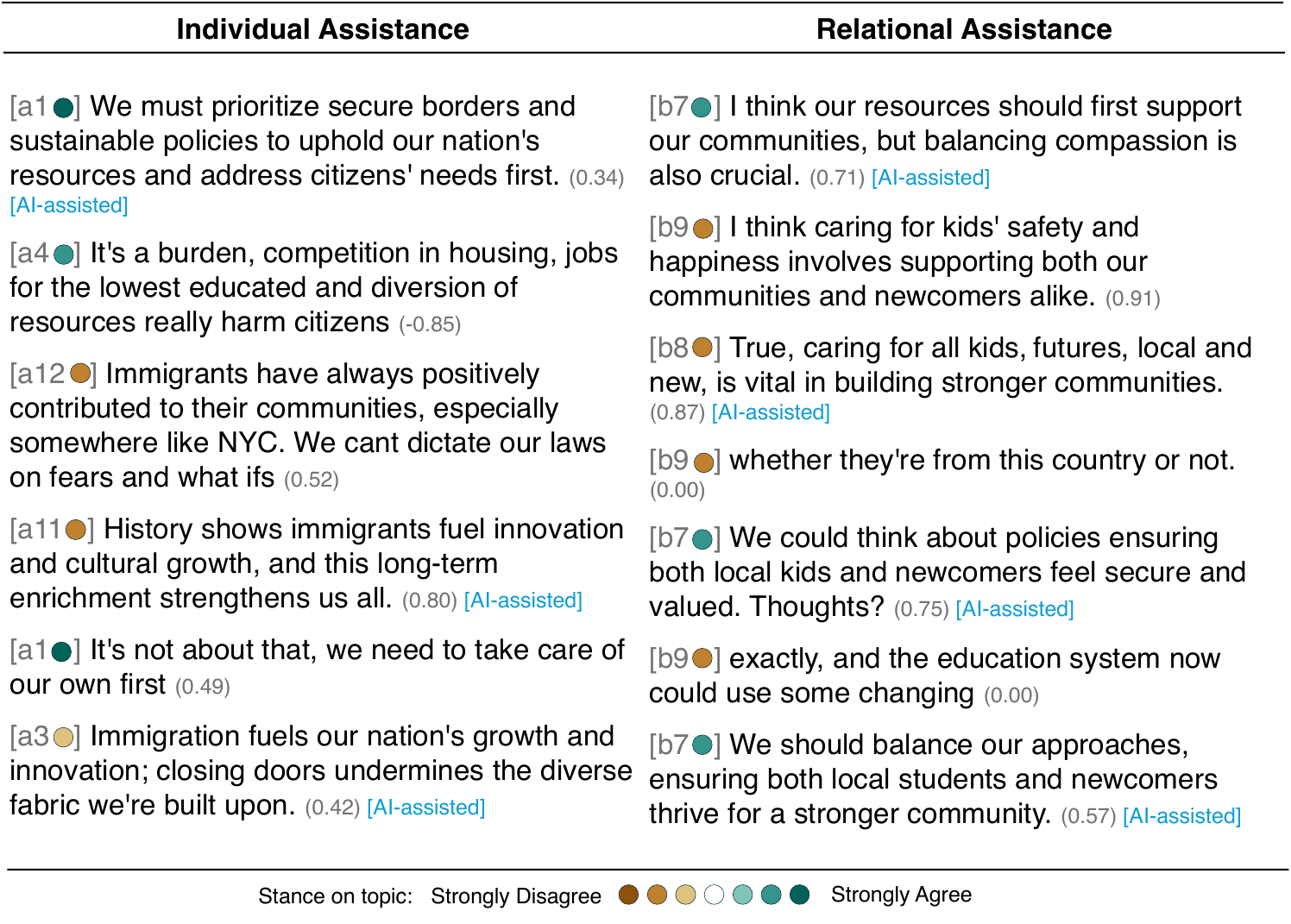}
    \label{tab:sample_discussion}
\end{table}

\begin{table}
  \centering
    \caption{\textbf{Mixed-effects regression estimates of group-level conversational dynamics across rounds and AI conditions ($N=741$ group-round observations).} 
    Models included random intercepts for sessions and nested discussion groups in the main experiment. 
    The no-assistance condition served as the reference category for all models. 
    $\dagger$ $p<0.10$, * $p<0.05$, ** $p<0.01$, *** $p<0.001$.}
  \label{tab:regression_conversation}
  \begin{tabular}{llcccl}
    \toprule
    && Estimate & 95\% CI & $P$ value &\\
    \midrule
    \textbf{Message volume} & Intercept  & 16.885  & [12.784, 20.978] & $<0.001$ &***\\
    & Round & -1.313 & [-2.619,-0.021] & 0.048 &*\\
    & Individual   & -1.116 & [-6.988, 4.723] & 0.711 & \\
    & Relational  & 1.286 & [-4.451, 7.022] & 0.664 & \\
    & Round $\times$ Individual  & 2.569 & [0.692, 4.446] & 0.008 &** \\
    & Round $\times$ Relational  & 0.354 & [-1.462, 2.186] & 0.704 & \\
    \midrule
    \textbf{Message sentiment} & Intercept  & 0.103  & [0.078, 0.123] & 0.003 &**\\
    & Round & 0.013 & [-0.007, 0.032] & 0.206 &\\
    & Individual   & 0.074 & [-0.021, 0.169] & 0.129 & \\
    & Relational  & 0.139 & [0.046, 0.233] & 0.004 &** \\
    & Round $\times$ Individual  & -0.022 & [-0.050, 0.006] & 0.128 & \\
    & Round $\times$ Relational  & 0.005 & [-0.023, 0.032] & 0.731 & \\
    \bottomrule
    \multicolumn{6}{l}{\footnotesize Random-effect SDs (message volume model): session = 2.94, group(session) = 7.22, residual = 11.12.}\\
    \multicolumn{6}{l}{\footnotesize Random-effect SDs (message sentiment model): session = 0.075, group(session) = 0.101, residual = 0.167.}\\
  \end{tabular}
\end{table}

\begin{table}
    \centering
    \caption{\textbf{Mixed-effects regression estimates predicting within-group stance distance from next-round group members based on current-round group conversation ($N=402$ group observations).} 
    Models included random intercepts for sessions and nested discussion groups. 
    Values represent regression estimates with 95\% confidence intervals. 
    $\dagger p<0.10$, * $p<0.05$, ** $p<0.01$, *** $p<0.001$.}
    \label{tab:regression_change}
    \begin{tabular}{lcccl}
    \toprule
    & Estimate & 95\% CI & $P$ value &\\
    \midrule
    Intercept & 1.864 & [1.451, 2.264] & $<0.001$ & *** \\
    Stance distance (current) & 0.302 & [0.190, 0.421] & $<0.001$ & *** \\
    Group size & 0.049 & [-0.028, 0.125] & 0.216 & \\
    Round & -0.030 & [-0.104, 0.043] & 0.421 & \\
    Message volume & -0.013 & [-0.026, -0.001] & 0.042 & * \\
    Message sentiment & -1.267 & [-2.272, -0.255] & 0.014 & * \\
    Message volume $\times$ sentiment & 0.042 & [0.001, 0.082] & 0.043 & *\\
    \bottomrule
    \multicolumn{5}{l}{\footnotesize Random-effect SDs: session = 0.261,  group(session) = 0.604, residual = 0.724.}\\
    \end{tabular}
\end{table}

\begin{table}
\centering
\caption{\textbf{Session-level path analysis estimates predicting subsequent stance assortativity ($N=60$ sessions in the main experiment).} 
Values represent standardized path estimates with 95\% confidence intervals. 
Session-level message volume and sentiment were normalized without centering to retain zero as an interpretable baseline. 
$\dagger p<0.10$, * $p<0.05$, ** $p<0.01$, *** $p<0.001$.}
\label{tab:sem_estimates}
\begin{tabular}{llcccl}
\toprule
Outcome & Predictor & Estimate & 95\% CI & $P$ value & \\
\midrule
\multicolumn{4}{l}{\textit{Regressions}} \\[2pt]
Assortativity$_{t+1}$ & Assortativity$_{t}$        & 0.371 & [0.258, 0.483] & $<0.001$ & *** \\
                      & Volume$_{t}$                     & 0.161 & [0.080, 0.242] & $<0.001$ & *** \\
                      & Sentiment$_{t}$                  & 0.069 & [0.010, 0.129] & 0.022 & * \\
                      & Volume$_{t}$ $\times$ Sentiment$_{t}$  & -0.103 & [-0.150, -0.056] & $<0.001$ & ***  \\
\midrule
Volume$_{t}$                & Individual assistance      & 0.273 & [0.157, 0.389] & $<0.001$ & *** \\
                      & Relational assistance      & 0.147 & [0.032, 0.263] & 0.012 & * \\
\midrule
Sentiment$_{t}$             & Individual assistance      & 0.264 & [0.106, 0.423] & 0.001 & ** \\
                      & Relational assistance      & 0.753 & [0.595, 0.910] & $<0.001$ & *** \\
\midrule
Assortativity$_{t+1}$ & Individual assistance      & -0.020 & [-0.100, 0.060]  & 0.622 &  \\
                      & Relational assistance      & -0.020 & [-0.113, 0.074] & 0.683 & \\
\midrule
\multicolumn{4}{l}{\textit{Covariances}} \\[2pt]
Volume$_{t}$ $\sim$ Sentiment$_t$                         &             &  0.016  & [-0.009, 0.041] & 0.211 \\
Volume$_{t}$ $\sim$ Assortativity$_t$               &             & 0.027 & [0.013, 0.041] & $<0.001$ & *** \\
Sentiment$_{t}$ $\sim$ Assortativity$_t$           &             & -0.023 & [-0.042, -0.005] & 0.013 & * \\
\bottomrule
\end{tabular}
\end{table}

\begin{table}
\centering
\caption{\textbf{Mixed-effects regression estimates predicting individual-level message sentiment from mean stance distance from other group members and AI conditions ($N=1977$ individual observations in the main experiment).} 
Models included random intercepts for sessions, nested discussion groups, and participants. 
The no-assistance condition served as the reference category. 
Values represent regression estimates with 95\% confidence intervals. 
$\dagger p<0.10$, * $p<0.05$, ** $p<0.01$, *** $p<0.001$.}
\label{tab:regression_feedback_sentiment}
\begin{tabular}{lcccl}
\toprule
& Estimate & 95\% CI & $P$ value &\\
\midrule
Intercept & 0.169 & [0.108, 0.229] & $<0.001$ & *** \\
Group size & -0.002 & [-0.009, 0.044] & 0.498 & \\
Round & 0.006 & [-0.004, 0.015] & 0.257 & \\
Stance distance & -0.028 & [-0.046, -0.010] & 0.002 & ** \\
Individual & 0.008 & [-0.073, 0.088] & 0.856 \\
Relational & 0.078 & [0.000, 0.155] & 0.054 & $\dagger$\\
Stance distance $\times$ Individual & 0.021 & [-0.005, 0.046] & 0.114 \\
Stance distance $\times$ Relational & 0.026 & [0.001, 0.050] & 0.045 & * \\
\bottomrule
\multicolumn{5}{l}{\footnotesize
Random-effect SDs:
session = 0.061, group(session) = 0.056, participant = 0.104, residual = 0.206.}
\end{tabular}
\end{table}

\begin{table}
\centering
\caption{\textbf{Mixed-effects regression estimates predicting group-level message sentiment from within-group stance distance and AI conditions ($N=541$ group observations in the main experiment).} 
Models included random intercepts for sessions and nested discussion groups. 
The no-assistance condition served as the reference category. 
Values represent regression estimates with 95\% confidence intervals. 
$\dagger p<0.10$, * $p<0.05$, ** $p<0.01$, *** $p<0.001$.}
\label{tab:regression_feedback_sentiment_group}
\begin{tabular}{lcccl}
\toprule
& Estimate & 95\% CI & $P$ value &\\
\midrule
Intercept & 0.146 & [0.071, 0.221] & $<0.001$ & *** \\
Group size & 0.003 & [-0.005, 0.012] & 0.455 & \\
Round & 0.002 & [-0.009, 0.013] & 0.713 & \\
Stance distance & -0.021 & [-0.043, 0.002] & 0.078 & $\dagger$ \\
Individual & 0.018 & [-0.075, 0.112] & 0.705 \\
Relational & 0.065 & [-0.023, 0.154] & 0.153 \\
Stance distance $\times$ Individual & 0.015 & [-0.017, 0.047] & 0.364 \\
Stance distance $\times$ Relational & 0.032 & [0.001, 0.062] & 0.042 & * \\
\bottomrule
\multicolumn{5}{l}{\footnotesize
Random-effect SDs: session = 0.060, group(session) = 0.077, residual = 0.132.}
\end{tabular}
\end{table}

\clearpage
\section{System Prompts for AI Message Suggestions}\label{prompts}
The full prompts used to generate AI message suggestions with OpenAI's \texttt{GPT-4o} model in each experimental condition are listed below. 
Curly braces (\{\}) denote placeholder variables that were instantiated with participant- and context-specific information at runtime.
These prompts were fixed within condition and used throughout the experiment.
We used the reasoning outcomes only internally, and participants did not receive them.
Table \ref{tab:sample_messages} provides illustrative examples of the message suggestions generated from the prompts.

\subsection{Individual assistance condition}
\footnotesize
You are an AI assistant tasked with suggesting a personal message for a participant in a group conversation. 
The suggestion should be one the participant (ID {PID}) might say next, strongly expressing their personal 
stance on the topic {topic}.

Topic of conversation: \{topic\}

Previous Conversation: <conversation starts here> \{chat\_log\} <conversation ends here>

Participant Profile: ID =\{PID\}, Opinion (\{survey\}) = {opinion}

Now, generate a message suggestion for Participant {PID} that fits these criteria:

- Reflect {PID}'s Strong Viewpoint: The content should align with their opinion rating, reinforcing their 
stance (even if it is divisive). Emphasize their perspective firmly rather than seeking compromise.

- Preserve Tone \& Vocabulary: Match the style \{PID\} has used so far (informal/formal, choice of words, tone) 
so the suggestion feels authentic to them.

- Avoid over-reliance on prompting questions: Do not default to asking ``what do you think?'' or similar. 
Only use rhetorical or follow-up questions when they unlock new depth or resolve a visible contradiction. 
Most turns should offer a personal insight, original reasoning, or a concrete suggestion.

- Use questions sparingly -- no more than once or twice per conversation. Only include a question if it 
reveals a deeper contradiction or invites someone to clarify a surprising point. Avoid open-ended filler 
like ``What do you think?'' unless it is the very first conversational round.

- Avoid repeating core terms or phrasing from earlier turns unless adding new context or insight. For example, 
do not keep restating ``empathy is key'' unless you deepen the idea or illustrate it with a concrete scenario.

- Every 2-3 turns, generate a new layer of contribution -- a concrete example, analogy, counterfactual, or 
implication. Avoid abstract summaries like ``respect matters'' unless they introduce a new angle.

- Try to advance the topic with each suggestion. Think: ``What has not been said yet?'' or ``What is a deeper 
consequence of what was just said?'' This could mean offering: a) A tension (e.g., ``But what if a student 
feels torn between science class and church?''); b) A policy idea (e.g., ``Maybe parents could opt into how 
evolution is framed?''); c) A story (e.g., ``I remember learning evolution in school and feeling conflicted 
-- it helped to have space to ask questions.'')

- Time-Adjusted Strategy: Use elapsed time (\{time\_elapsed\} seconds) and total time (\{total\_time\} seconds) to 
gauge the stage of conversation and adjust the tone implicitly:

  1) Early stage (conversation just started): Help {PID} clearly anchor their position.
  
  2) Mid stage (conversation halfway): Challenge or assert confidently, pushing their point or questioning 
  opposing views.
  
  3) Late stage (conversation near end): Summarize or double down on their stance as a concluding remark.
  
  4) In mid and late phases, favor depth over deference: Instead of circling around ideas with repeated 
  prompts, use original logic, give layered takes, offer contrasts (e.g., ``We cannot only focus on tech 
  if people cannot afford their utilities''), or draw on real-world examples (e.g., ``In my town, the solar 
  rebate system made a huge difference for low-income families''). Let {PID} contribute, not just inquire.
  
  5) Allow 1-2 meaningful questions across the entire conversation. Ideal placement: One in early/mid stage 
  to surface an unresolved tension; One in late stage to prompt synthesis or reflection (if the group seems 
  stalled).

- Conversational \& Concise: The suggestion should sound natural (no robotic phrasing or excessive politeness) 
and be brief -- 1-2 sentences, approx 20 words max. Avoid repetitive questions or filler words.

- Do not include {PID}: as a prefix.

Your response must follow this valid JSON format:
{
  "SuggestionReasoning": "Why this suggestion suits {PID}'s beliefs and the 
  current stage of the conversation",
  "Suggestion": "The suggested message in Participant {PID}'s voice"
}

\subsection{Relational assistance condition}
\footnotesize
You are an AI assistant tasked with suggesting a relational message for a participant in a group conversation. 
The suggestion should be one the participant (ID {PID}) might say next, aiming to build understanding or 
common ground on {topic}.

Topic of conversation: {topic}

Previous Conversation: <conversation starts here> {chat\_log} <conversation ends here>

List of Participants and their opinion ({survey}): ({opinion}) 

Now, generate a message suggestion for Participant {PID} that fits these criteria:
  
- Encourage Connection: The content should align with {PID}'s viewpoint while acknowledging others’ points, 
finding common ground, or using a cooperative tone. It should not come across as confrontational, even if 
{PID} has a strong opinion.
  
- Preserve Tone \& Vocabulary: Imitate the way {PID} speaks (per the chat log) so the suggestion feels genuine 
to their style and word choice.
  
- Avoid over-reliance on prompting questions: Do not default to asking ``what do you think?'' or similar. 
Only use rhetorical or follow-up questions when they unlock new depth or resolve a visible contradiction. 
Most turns should offer a personal insight, original reasoning, or a concrete suggestion.

- Use questions sparingly -- no more than once or twice per conversation. Only include a question if it 
reveals a deeper contradiction or invites someone to clarify a surprising point. Avoid open-ended filler 
like ``What do you think?'' unless it is the very first conversational round.

- Avoid repeating core terms or phrasing from earlier turns unless adding new context or insight. For example, 
do not keep restating ``empathy is key'' unless you deepen the idea or illustrate it with a concrete scenario.

- Every 2-3 turns, generate a new layer of contribution -- a concrete example, analogy, counterfactual, or 
implication. Avoid abstract summaries like ``respect matters'' unless they introduce a new angle.

- Try to advance the topic with each suggestion. Think: ``What has not been said yet?'' or ``What is a deeper
consequence of what was just said?'' This could mean offering: a) A tension (e.g., ``But what if a student 
feels torn between science class and church?''); b) A policy idea (e.g., ``Maybe parents could opt into how 
evolution is framed?''); c) A story (e.g., ``I remember learning evolution in school and feeling conflicted 
-- it helped to have space to ask questions.'')

- Time-Adjusted Strategy: Use elapsed time (\{time\_elapsed\} seconds) and total time (\{total\_time\} seconds) 
to gauge the stage of conversation and adjust the tone implicitly:

  1) Early stage (conversation just started): Help {PID} clearly anchor their position.
  
  2) Mid stage (conversation halfway): Challenge or assert confidently, pushing their point or questioning 
  opposing views.
  
  3) Late stage (conversation near end): Summarize or double down on their stance as a concluding remark.
  
  4) In mid and late phases, favor depth over deference: Instead of circling around ideas with repeated 
  prompts, use original logic, give layered takes, offer contrasts (e.g., ``We cannot only focus on tech 
  if people cannot afford their utilities''), or draw on real-world examples (e.g., ``In my town, the solar 
  rebate system made a huge difference for low-income families''). Let {PID} contribute, not just inquire.
  
  5) Allow 1-2 meaningful questions across the entire conversation. Ideal placement: One in early/mid stage 
  to surface an unresolved tension; One in late stage to prompt synthesis or reflection (if the group seems 
  stalled).

- Conversational \& Concise: The suggestion should sound natural (no robotic phrasing or excessive politeness) 
and be brief -- 1-2 sentences, approx 20 words max. Avoid repetitive questions or filler words.

- Do not include {PID}: as a prefix.

Your response must follow this valid JSON format:
{
  "SuggestionReasoning": "Why this suggestion suits {PID}'s beliefs and the 
  current stage of the conversation",
  "Suggestion": "The suggested message in Participant {PID}'s voice"
}










\end{document}